\documentclass[aps,prx, reprint]{revtex4-2}

\usepackage{graphicx}
\usepackage{dcolumn}
\usepackage{amsfonts}
\usepackage{braket}
\usepackage{multirow}
\usepackage{threeparttable}
\usepackage{xspace}
\usepackage{verbatim}
\usepackage[version=4]{mhchem}
\usepackage{siunitx}
\usepackage{qcircuit}
\usepackage{lipsum}
\usepackage{amsmath}
\usepackage{mathtools}
\usepackage{xcolor}
\usepackage{xspace}
\usepackage{ifthen}

\usepackage[a-1b]{pdfx}
\usepackage{hyperref}

\newcommand*{\Eh}{$E_{\rm h}$\xspace}

\providecommand{\norm}[1]{| #1 |^{2}}
\providecommand{\normnorm}[1]{\lVert#1\rVert}

\newcommand{\cop}[1]{\hat{a}^{\dagger}_{#1}}
\newcommand{\aop}[1]{\hat{a}_{#1}}

\newcommand*{\Ud}{\hat{U}^{\dagger}}

\newcommand*{\Uk}{\hat{U}^{(k)}}
\newcommand*{\Ukd}{\hat{U}^{(k)\dagger}}

\newcommand*{\qucc}{dUCC\xspace}

\providecommand{\norm}[1]{\lVert#1\rVert}

\begin{document}

\title{
Simulating Many-Body Systems with a Projective Quantum Eigensolver
}
\author{Nicholas H. Stair}
\email{nstair@emory.edu}
\author{Francesco A. Evangelista}
\email{francesco.evangelista@emory.edu}
\affiliation{Department of Chemistry and Cherry Emerson Center for Scientific Computation, Emory University, Atlanta, GA 30322, USA}
\date{\today}

\begin{abstract}
We present a new hybrid quantum-classical algorithm for optimizing unitary coupled-cluster (UCC) wave functions deemed the projective quantum eigensolver (PQE), amenable to near-term noisy quantum hardware.
Contrary to variational quantum algorithms, PQE optimizes a trial state using  residuals (projections of the Schr\"{o}dinger equation) rather than energy gradients.
We show that the residuals may be evaluated by simply measuring two energy expectation values per element.
We also introduce a selected variant of PQE (SPQE) that uses an adaptive ansatz built from arbitrary-order particle-hole operators, offering an alternative to gradient-based selection procedures.
PQE and SPQE are tested on a set of molecular systems covering both the weak and strong correlation regimes, including hydrogen clusters with 4--10 atoms and the \ce{BeH2} molecule.
When employing a fixed ansatz, we find that PQE can converge disentangled (factorized) UCC wave functions to essentially identical energies as variational optimization while requiring fewer computational resources.
A comparison of SPQE and adaptive variational quantum algorithms shows that---for ans\"{a}tze containing the same number of parameters---the two methods yield results of comparable accuracy.
Finally, we show that SPQE performs similar to, and in some cases, better than selected configuration interaction and the density matrix renormalization group on 1--3 dimensional strongly correlated \ce{H10} systems in terms of energy accuracy for a given number of variational parameters.

\end{abstract}

\maketitle

\section{Introduction}
Efficient quantum algorithms for determining the ground and excited states of many-body systems are of fundamental interest to chemistry, condensed matter physics, and materials science \cite{Lloyd:1996ch, aspuru2005simulated, kandala2017hardware, kandala2019error}.
The ability of quantum devices to represent $N$-body states with qubits scaling linearly in $N$ make them particularly appealing for representing highly entangled states, as is common in systems with strongly correlated electrons.
Therefore, quantum (and hybrid quantum-classical) algorithms offer an alternative to methods such as the density matrix renormalization group \cite{White1992DensityMatrix} (DMRG), selected configuration interaction \cite{Huron1973IterativePerturbation, Buenker1974IndividualizedConfiguration} (SCI),  determinant-based Monte-Carlo \cite{Booth2009FermionMonte}, variants of coupled-cluster (CC) theory \cite{coester1960short, vcivzek1966correlation} amenable to treating strong correlation \cite{Piecuch:1990wa,piecuch2005renormalized, Limacher2013NewMean, Bulik2015CanSingle}, and multireference CC (MRCC) methods \cite{vcivzek1969use, lindgren1978coupled, jeziorski1981coupled,Lyakh:2012cn,Kohn:2013cp,evangelista2018perspective}.
Although these classical algorithms can accurately predicted energies and properties of certain classes of strongly correlated systems, they still have high-order polynomial or exponential cost in the general case.

Since Feynman's proposal to use a controlled quantum system to carry out simulations \cite{feynman1982simulating}, significant algorithmic and experimental advances have been made.
The earliest demonstrations of quantum simulation for small molecules \cite{aspuru2005simulated} utilized the quantum phase estimation algorithm \cite{kitaev1995quantum, Abrams:1997ha, Abrams:1999ur}  with Suzuki--Trotter decomposed time evolution \cite{trotter1959product, suzuki1993improved} of an adiabatically prepared trial state.
It is believed that some combination of these techniques will permit the efficient simulation \cite{von2020quantum, lee2020even, babbush2018low, reiher2017elucidating} of certain classes of Hamiltonians \cite{kempe2006complexity}, but that they may require deep circuits with high fidelity, a requirement incompatible with current noisy intermediate-scale quantum (NISQ) devices \cite{preskill2018quantum}.
Several low-depth quantum-classical hybrid algorithms have been developed for NISQ hardware.
These algorithms prepare and measure properties of many-body states on a quantum device, but they store and optimize the parameters that define such states on a classical computer.
The variational quantum eigensolver (VQE) approach \cite{Peruzzo:2014kca, yung2014transistor, McClean:2016bs, grimsley2019adaptive} has been used in several landmark experiments, demonstrating quantum calculations on non-trivial molecular systems \cite{OMalley:2016dc, kandala2017hardware, colless2018computation, shen2017quantum, hempel2018quantum, nam2020ground}.
In VQE, the ground state is approximated by a normalized trial state $\ket{\tilde{\Psi}} = \hat{U}(\mathbf{t}) \ket{\Phi_0}$, in which the unitary operator $\hat{U}(\mathbf{t})$ depends on the parameter vector $\mathbf{t}$ and $\Phi_0$ is (usually) an unentangled reference state.
The VQE energy ($E_{\rm{VQE}}$) is then obtained by minimization of the trial state energy expectation value as
\begin{equation}
\label{eq:vqe}
E_{\rm{VQE}} = \min_\mathbf{t} \bra{\Phi_0} \hat{U}^\dagger(\mathbf{t}) \hat{H} \hat{U}(\mathbf{t}) \ket{\Phi_0}.
\end{equation}
The VQE scheme employs an optimization algorithm running on a classical computer to minimize the energy expectation value, with all inputs (energy/gradients) being evaluated with the help of a quantum computer.
An important advantage of VQE over classical many-body methods is the ability to use trial states that cannot be represented efficiently on a classical computer.
VQE was initially implemented with an exponential operator ansatz inspired by unitary coupled-cluster (UCC) theory \cite{szalay1995alternative, taube2006new, cooper2010benchmark, evangelista2011alternative, harsha2018difference,Filip:2020ib,Chen:2021fa}, but has more recently been extended to hardware-efficient \cite{kandala2017hardware} and qubit-space \cite{Ryabinkin:2018jw} UCC variants as well.
We exclusively use the abbreviation UCC to refer to unitary coupled-cluster theory, and not unrestricted formulations of conventional coupled-cluster methods \cite{knowles1993coupled}, which historically share this abbreviation.
Other promising hybrid approaches include quantum imaginary time evolution \cite{motta2019determining,PRXQuantum.2.010317}, and quantum subspace diagonalization techniques \cite{mcclean2017hybrid, motta2019determining, Parrish:2019tc, Stair_2020, huggins2020non}.

Despite the indisputable importance of VQE in the field of quantum simulation, there are a few drawbacks that make its practical application challenging to large-scale problems.
One such issue is the slow convergence of VQE due to noise in the measured energy and gradients, and the large-scale nonlinear nature of the optimization problem.
These issues are compounded by the sizable number of total measurements needed for operator averaging \cite{wecker2015progress}.
Another challenge is the potentially large number of classical parameters and resulting circuit depth necessary to predict sufficiently accurate energies.
These two problems are likely exacerbated in systems with strongly correlated electrons.

Progress addressing these deficiencies of VQE has been made on several fronts.
For example, grouping commuting Pauli operators \cite{McClean:2016bs, kandala2017hardware, gokhale2019minimizing, yen2020measuring, verteletskyi2020measurement}, utilizing integral factorization strategies \cite{huggins2021efficient}, and employing alternative bases~\cite{babbush2018low, mcclean2020discontinuous} have been shown to reduce the number of measurements needed for operator averaging.
Concurrently, computationally feasible approaches for measuring analytical gradients with quantum devices using the parameter-shift rule \cite{schuld2019evaluating}, or its recent lower-cost variant \cite{kottmann2020feasible}, have allowed gradient-based VQE to become potentially realizable on NISQ hardware.
Other advances, of particular importance to this work, include VQE ans\"{a}tze constructed iteratively, as done in the adaptive derivative-assembled pseudo-Trotterized VQE \cite{grimsley2019adaptive} (ADAPT-VQE) and the iterative qubit coupled-cluster \cite{ryabinkin2020iterative} (iQCC) methods.
The primary advantage of ADAPT-VQE and iQCC is their ability to produce compact ans\"{a}tze that result in fewer classical parameters, and shallower quantum circuits than those from UCC truncated to a given particle-hole rank.
However, these advantages come at the cost of a greater number of energy and gradient evaluations for optimizing and selecting new unitary operators.
Investigating more efficient ways to select important operators is an ongoing area of research \cite{zhang2020mutual, liu2020efficient}.

In this work, we present an alternative to VQE for optimizing the amplitudes of a factorized form of the UCC ansatz (often referred to as Trotterized \cite{McClean:2016bs} or quantum \cite{Barkoutsos:2018hm} UCC), given by a product of exponential operators rather than the exponential of a sum of operators. We refer to this ansatz as disentangled UCC (\qucc)---a terminology borrowed from the field of Lie theory---to reflect the fact that it is not an approximation of UCC \cite{evangelista2019exact}.
Inspired by the projective approach used in classical coupled-cluster theory \cite{vcivzek1966correlation, vcivzek1969use}, we propose an alternative trial state optimization algorithm that we deem the projective quantum eigensolver (PQE).
PQE does not rely on variational minimization and therefore does not require evaluation of the energy gradients.
Instead, PQE requires only the evaluation of residuals, that is, projections of the Schr\"{o}dinger equations onto a linearly independent basis.
As shown in this paper, residuals may be easily measured on NISQ devices with similar or fewer measurements than analytical gradients, and require quantum circuits that contain only one additional exponential term.
We also propose a new selection scheme for identifying important operators based on the residual vector.
This selected variant of PQE (SPQE) requires no pre-defined operator pool and employs only a small number of measurements to identify important operators.
To demonstrate the practical advantages of PQE, we perform a comparison of VQE and PQE using a fixed \qucc  ansatz for several molecular systems in the regime of weak and strong correlation, also considering the effect of stochastic noise.
Finally, we compare SPQE against the ADAPT-VQE approach, selected configuration interaction, and the density matrix renormalization group.

\section{Theory}
\label{sec:theory}

\subsection{The Projective Quantum Eigensolver approach}
\label{sec:pqe_theory}

In this work, we propose to obtain the ground state of a general many-body system using a projective approach.
Like in VQE, we approximate the ground state using a trial state $\ket{\tilde{\Psi}(\mathbf{t})}= \hat{U}(\mathbf{t}) \ket{\Phi_0}$.
After inserting the definition of the trial state in the Schr\"{o}dinger equation and left-multiplying by $\hat{U}^\dagger(\mathbf{t})$, we obtain the condition
\begin{equation}
\hat{U}^\dagger(\mathbf{t}) \hat{H} \hat{U}(\mathbf{t}) \ket{\Phi_0} = E \ket{\Phi_0}.
\end{equation}
Projection onto the reference state $\Phi_0$ yields the PQE energy ($E_\text{PQE}$)
\begin{equation}
E_\text{PQE}(\mathbf{t}) = \bra{\Phi_0} \hat{U}^\dagger(\mathbf{t}) \hat{H} \hat{U}(\mathbf{t}) \ket{\Phi_0},
\label{eq:ucc1}
\end{equation}
a quantity that is still an upper bound to the exact ground state energy.
Projections onto the complete set of orthonormal many-body basis functions complementary to $\Phi_0$, here denoted as $Q = \{\Phi_\mu \}$, yields a set of residual conditions
\begin{equation}
r_\mu(\mathbf{t}) \equiv \bra{\Phi_\mu} \hat{U}^\dagger(\mathbf{t}) \hat{H} \hat{U}(\mathbf{t}) \ket{\Phi_0} = 0 \quad \forall \Phi_\mu \in Q,
\label{eq:ucc2}
\end{equation}
where $r_\mu$ is an element of the residual vector and $\mu$ runs over all elements of the many-body basis.
Eqs.~\eqref{eq:ucc1} and \eqref{eq:ucc2} form a system of nonlinear equations in the parameter vector $\mathbf{t}$, that may be solved via a classical iterative solver.
For an approximate ansatz with number of parameters less than the dimension of the $Q$ space, Eq.~\eqref{eq:ucc2} can be enforced only for a subset of the residuals.
Then, the complete projection space $Q$ can be partitioned into two sets: i) $R$, the space of basis functions $\Phi_\mu$ for which $r_\mu = 0$ is enforced, and ii) $S = Q \setminus R$ the complementary space for which $r_\mu$ may not be null.

\begin{figure}[ht!]
\centering
\includegraphics[width=3.375in]{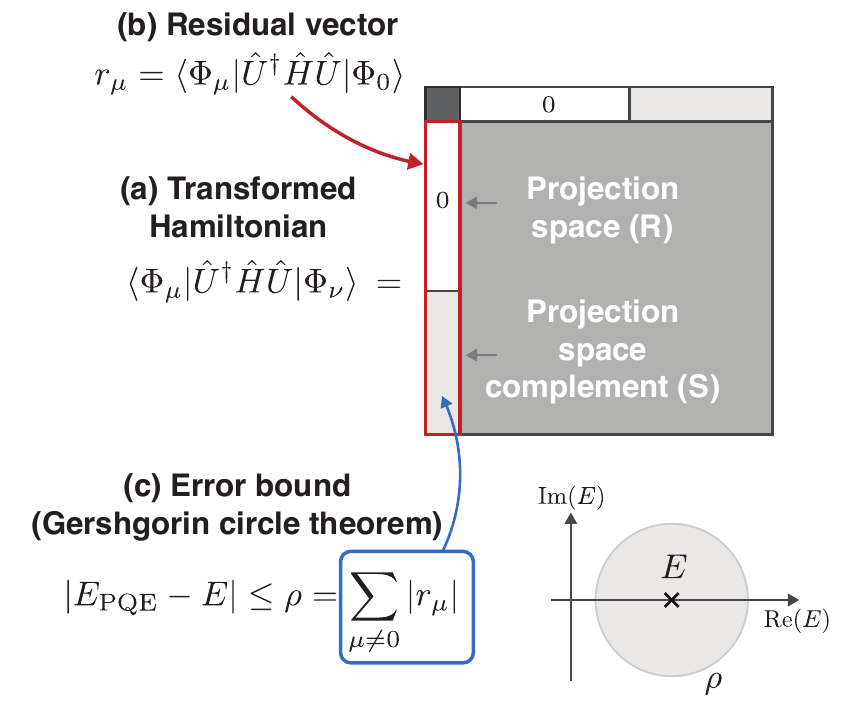}
\caption{Connection between the norm of the PQE residual and the energy error via the Gershgorin circle theorem. (a) Structure of the transformed Hamiltonian in the basis of orthogonal states $\{\Phi_\mu\}$. (b) The residual vector is the first column of the transformed Hamiltonian matrix (first element excluded).
(c) The difference between the approximate ground-state PQE energy ($E_{\mathrm{PQE}}$) and the exact eigenvalue ($E$) is bound by the radius $\rho$, which is equal to the 1-norm of the residual vector.
The part of $r_\mu$ that corresponds to states in the projection manifold $R$ is null for the PQE solution, while elements involving projections on the space $S = Q \setminus R$ is generally nonzero and determines the value of $\rho$.}
\label{fig:residual}
\end{figure}

Figure~\ref{fig:residual} illustrates the connection between the PQE residual condition and the uncertainty in the ground-state energy estimated via Eq.~\eqref{eq:ucc1}.
By the Gershgorin circle theorem, the difference between the exact  ($E$) and the PQE ($E_{\mathrm{PQE}}$) ground-state energy, $|E_{\mathrm{PQE}} - E|$, is bound by the radius $\rho = \sum_{\mu \neq 0} |r_\mu|$, where $\mu$ runs over the entire many-body basis, excluding the reference determinant.
Therefore, when the residual is null ($\rho = 0$), the PQE energy is exact.
When the PQE equation is satisfied only by a subset of the many-body basis---as in the case of an approximate trial state---the error $|E_{\mathrm{PQE}} - E|$ is bound by the sum of the absolute value of the residual elements $|r_\nu|$ with $\Phi_\nu \in S$, for which the PQE equation is not satisfied.

Note that the residual condition [Eq.~\eqref{eq:ucc2}] is satisfied by any eigenstate, and the Gershgorin circle theorem error bound applies also to excited states.
A potential disadvantage is that PQE could converge on an excited state (an issue we did not experience in this study).
However, this feature could be used to formulate excited state algorithms based on PQE, which use the residual condition as a criterion for convergence and do not require costly measurement of the variance, as is commonly done in VQE \cite{McClean:2016bs}.

The PQE is a general approach, however, in the following, we will focus on its applications to interacting fermions using a disentangled (factorized) form of the unitary coupled-cluster ansatz.
We assume that our system is described by the general two-body Hamiltonian
\begin{equation}
\hat{H} = \sum_{pq} h_{pq} \cop{p} \aop{q}
+ \frac{1}{4} \sum_{pqrs}
v_{pqrs} \cop{p} \cop{q} \aop{s} \aop{r},
\end{equation}
where $\aop{p}$ ($\cop{q}$) is a fermionic annihilation (creation) operator, while $h_{pq}$ and $v_{pqrs}$
are one-electron and antisymmetrized two-electron integrals, respectively \cite{crawford2000introduction}.

\subsection{Traditional and disentangled unitary coupled-cluster ans\"{a}tze}
\label{sec:trad_dis_ucc}

In UCC, the reference state is an easily-prepared single determinant $\ket{\Phi_0} = \ket{\psi_1 \psi_2 \cdots}$ specified by the occupied spin orbitals $\{ \psi_i \}$.
 A UCC unitary is parameterized using a pool
of anti-Hermitian operators $\mathcal{P} = \{  \hat{\kappa}_\mu : \mu =1 ,\ldots, N_\mathrm{op}^\mathrm{pool} \}$.
A generic anti-Hermitian operator $\hat{\kappa}_\mu = \hat{\tau}_\mu - \hat{\tau}_\mu^\dagger$ is defined in terms of the particle-hole excitation operators
$ \hat{\tau}_\mu \equiv  \hat{\tau}_{ij\cdots}^{ab\cdots} = \cop{a} \cop{b} \cdots \aop{j} \aop{i}$.
Note that we have re-interpreted $\mu$ as the multi-index $\mu \equiv ((i,j,..),(a,b,..))$ of unique excitations from hole/occupied ($\psi_i \psi_j \cdots$) to particle/unoccupied ($\psi_a \psi_b \cdots$) spin orbitals.
Using this parameterization, when a cluster operator $ \hat{\kappa}_\mu$ acts on the reference, it generates elements of the many-body basis (excited determinants) of the form
\begin{equation}
\ket{\Phi_\mu} = \hat{\kappa}_\mu \ket{\Phi_0} = \ket{\Phi_{ij\cdots}^{ab\cdots}},
\end{equation}
and since in the case of a UCC (or dUCC) ansatz there is a 1-to-1 correspondence between operators and determinants, we may label them with the same index.
Note that this operator basis satisfies the orthonormality condition $\bra{\Phi_0} \hat{\kappa}^\dagger_\mu  \hat{\kappa}_\nu \ket{\Phi_0} = \braket{\Phi_\mu|\Phi_\nu} = \delta_{\mu\nu}$.

In traditional UCC \cite{szalay1995alternative, taube2006new, cooper2010benchmark, evangelista2011alternative, harsha2018difference}, the wave function is generated by an exponential operator
\begin{equation}
\label{eq:trad_ucc}
\hat{U}(\mathbf{t}) = e^{\hat{\sigma}} = e^{\sum_\mu t_\mu \hat{\kappa}_\mu},
\end{equation}
assuming the cluster amplitudes $t_\mu$ are real.
In principle it is possible to construct a circuit that exactly implements the action of the UCC operator defined in Eq.~\eqref{eq:trad_ucc}, but in practice it is common to use a unitary with a simpler, and shallower, circuit.
This is frequently accomplished using a factorized (disentangled) form of the UCC ansatz
\begin{equation}
\label{eq:st_ucc}
\hat{U}(\mathbf{t})=
 \prod_\mu e^{ t_\mu \hat{\kappa}_\mu}.
\end{equation}
Because the operators $\hat{\kappa}_\mu$ do not commute, an ansatz of the disentangled form is uniquely defined by an \textit{ordered} set (or subset) of operators $\mathcal{A} = ( \hat{\kappa}_{\mu_i}: i = 1, \ldots, N_\mathrm{op} )$ built from the pool $\mathcal{P}$.
The operators in $\mathcal{A}$ are then used to form an ordered product of exponential unitaries
\begin{equation}
\label{eq:qucc}
\hat{U}(\mathbf{t})
= e^{t_{\mu_1} \hat{\kappa}_{\mu_1}}  e^{t_{\mu_2} \hat{\kappa}_{\mu_2}} \cdots e^{t_{\mu_{N_\mathrm{op}}} \hat{\kappa}_{\mu_{N_\mathrm{op}}}},
\end{equation}
where $t_{\mu_i}$ is the amplitude corresponding to the operator $\hat{\kappa}_{\mu_i}$.

The disentangled UCC ansatz may be viewed as a first-order Suzuki--Trotter approximation to UCC; however, it was recently shown \cite{evangelista2019exact} that dUCC is substantially different from conventional UCC \cite{szalay1995alternative, taube2006new, cooper2010benchmark, evangelista2011alternative, harsha2018difference}.
An arbitrary quantum state can always be represented in the form of Eq.~\eqref{eq:qucc} using particle-hole excitation/de-excitation operators \cite{evangelista2019exact}.
However, only certain operator orderings $\mathcal{A}$ of a complete operator pool (containing up to $N$-body particle/hole excitations) can represent any quantum state.
Nevertheless, orderings that do not satisfy this condition can exactly represent states that are close to the reference.
In this work we assume that all operators appear at most once in $\mathcal{A}$, but the more general case in which $\mathcal{A}$ contains repetitions has also been considered in other contexts \cite{grimsley2019adaptive, lee2018generalized}.

For PQE formulated using a UCC or dUCC trial state, it is possible to show (see Appendix~\ref{sec:appD}) that for an exact ansatz the residual condition [Eq.~\eqref{eq:ucc2}] and the VQE energy stationarity condition ($\partial E_\mathrm{VQE} / \partial t_\mu = 0$) are equivalent.
However, for an approximate ansatz we find that the gradient of the PQE energy contains a contribution due to the nonzero residual elements corresponding to the subspace $S$:
\begin{equation}
\label{eq:grad_residual_connection}
\frac{\partial E_\mathrm{PQE}(\mathbf{t})}{\partial t_\mu}
= 2 \, \mathrm{Re} \sum_{\Phi_\nu \in S}
r_\nu^*
\bra{\Phi_\nu} \hat{U}^\dagger(\mathbf{t}) \frac{\partial \hat{U}(\mathbf{t})}{\partial t_\mu}  \ket{\Phi_0},
\end{equation}
where $\mathbf{t}$ is a solution of the PQE equations in the projection space $R$.
Suppose that $R$ is chosen to be the space of single and double excitations and $S$ its complement.
Then this result shows that even if $r_\mu = 0$ for all singles and doubles, the gradient of $E_\mathrm{PQE}$ with respect to singles and doubles may not be zero because residuals $r_\nu$ corresponding to triple and higher excitations and the term $\bra{\Phi_0} \hat{U}^\dagger(\mathbf{t}) \partial \hat{U}(\mathbf{t})/\partial t_\mu
\ket{\Phi_\nu}$ are generally not null. Therefore, PQE and VQE energies obtained from approximate ans\"{a}tze will be different.

We note that the combination of PQE and dUCC produces energies that are additive for non-interacting fragments (size consistent) when using a localized basis.
This property follows from the fact that dUCC excitation operators for non-interacting fragments act only on orbitals localized on each fragment, and therefore, commute. Consequently, the dUCC wave function is multiplicatively separable (as long as the order of the operators within a fragment is preserved).
We have verified numerically that dUCC with singles and doubles trial states optimized with PQE energy are size consistent by performing calculations on H$_4$ + H$_2$ separated at a 1000 \AA{} and verified to within numerical convergence ($10^{-10}$ \Eh) that the energy is additive in the fragments.

\subsection{Numerical solution of the \qucc-PQE amplitude equation}

To realize the PQE scheme on a quantum computer, the reference state, the Hamiltonian, and the unitary must be represented in a qubit basis via a fermionic mapping.
After such a transformation, the Hamiltonian is a sum of the form $\hat{H} = \sum_\ell h_\ell \hat{O}_\ell,$ where $h_\ell$ is an electron integral multiplied by a coefficient and $\hat{O}_\ell = \prod_{i} \hat{\sigma}_{j_{\ell_i}}^{q_{\ell_i}}$ is a unique product of $\hat{\sigma}_{x}$, $\hat{\sigma}_{y}$, or $\hat{\sigma}_{z}$  Pauli operators acting on qubits $q_{\ell_i}$.
Similarly, each term in the unitary $\exp(t_\mu \hat{\kappa}_\mu )$ may be implemented using a combination of one- and two-qubit operators following standard approaches \cite{Peruzzo:2014kca, yung2014transistor, McClean:2016bs}.

To solve the PQE equations we measure the residuals corresponding to the operators contained in $\mathcal{A}$ on a quantum computer and update the parameter vector using a simple quasi-Newton iteration approach
\begin{equation}
\label{eq:fixed_point}
t_\mu^{(n +1)} = t_\mu^{(n)} + \frac{r^{(n)}_\mu}{\Delta_\mu},
\end{equation}
where the superscript ``$(n)$'' indicates the amplitude at iteration $n$.
The quantities $\Delta_\mu$ are standard M{\o}ller--Plesset denominators $\Delta_\mu \equiv \Delta_{ij\cdots}^{ab\cdots} = \epsilon_i + \epsilon_j + \ldots -\epsilon_a -\epsilon_b \ldots$ where $\epsilon_i$ are Hartree--Fock orbital energies.
This update equation is derived in Appendix~\ref{sec:appC} using Newton's method and taking the leading contributions to the Jacobian to be the diagonal elements of the Fock operator \cite{doi:https://doi.org/10.1002/9781119019572.ch13}.
It is further assumed that the amplitudes are small, so that the Jacobian can be approximated by terms linear in the operators $\hat{\kappa}_{\mu}$ and issues with non-commuting operators are avoided.
Therefore, convergence of this quasi-Newton scheme is not mathematically guaranteed if one or more amplitudes are large.
We found it useful to improve numerical stability and speed up convergence of the PQE equations, to combine amplitude updates via Eq.~\eqref{eq:fixed_point} with the direct inversion of the iterative subspace (DIIS) convergence accelerator algorithm \cite{pulay1980convergence,SCUSERIA1986236}.

It is important to note that the current formulation of PQE is compatible with any ansatz such that the metric matrix
\begin{equation}
S_{\mu_i \mu_j}   = \bra{\Phi_0} \hat{\kappa}^\dagger_{\mu_i}  \hat{\kappa}_{\mu_j} \ket{\Phi_0}, \quad \forall \hat{\kappa}_{\mu_i}, \hat{\kappa}_{\mu_j} \in \mathcal{A}
\end{equation}
is the identity.
In more general cases (e.g., when $S_{\mu_i \mu_j}$ is non-diagonal or singular), the PQE formalism requires a generalization of the amplitude update equations or the use of residual norm minimization instead of Eq.~\eqref{eq:ucc2}.
These variants of PQE would allow one to consider ans\"{a}tze that contain repeated operators in $\mathcal{A}$, employ general many-body operators,  or a basis of qubit operators.
These extensions go beyond the scope of this work and will be considered in future studies.

There are two advantages to the combination of the PQE and \qucc described  above. Firstly, as we will show in the following subsection, one element of the residual vector ($r_\mu$) can be evaluated with essentially the same resources  required to measure the energy in VQE.
Secondly, the magnitude of the residuals provides an indication of the importance of an excitation operator $\hat{\kappa}_\mu$, which in turn may be used to define a selection procedure to form the sequence of unitiaries that enter in $\hat{U}(\mathbf{t})$.
The next two subsections describe these two points in detail.

\subsection{Efficient measurement of the residual elements}
For a trial state built from the ordered pool $\mathcal{A}$, the number of the residual elements that must be evaluated to solve the PQE equations is equal to the size of the pool $|\mathcal{A}|$.
The PQE residuals can be expressed as the off-diagonal matrix elements of the operator $\bar{H} = \hat{U}^\dagger(\mathbf{t}) \hat{H} \hat{U}(\mathbf{t})$ as $r_\mu = \bra{\Phi_\mu} \bar{H} \ket{\Phi_0}$ (we use this notation throughout the paper, but we note that we never explicitly form the operator $\bar{H}$ on a classical computer).
Then, in principle, the residuals can be measured on a quantum computer using a variant of the Hadamard test \cite{aharonov2009polynomial}, but we have found an ancilla-free procedure in which these matrix elements are computed by measuring diagonal quantities.
Acting on the reference with the operator $e^{\theta \hat{\kappa}_\mu}$ yields the state
\begin{equation}
\ket{\Omega_\mu(\theta)} = e^{\theta \hat{\kappa}_\mu} \ket{\Phi_0} = \cos(\theta) \ket{\Phi_0} + \sin(\theta) \ket{\Phi_\mu},
\end{equation}
noting that the above expression is valid because $\hat{\kappa}_\mu\ket{\Phi_0} = \ket{\Phi_\mu}$ and $\hat{\kappa}^2_\mu\ket{\Phi_0} = -\ket{\Phi_0}$ (see also \cite{Filip:2020ib,Chen:2021fa}).
Taking the expectation value of the similarity transformed Hamiltonian with respect to $\Omega_\mu(\theta)$ using $\theta = \pi / 4$, and noting that the wave function is real, leads to the following equation for the residual elements
\begin{equation}
r_\mu = \bra{\Omega_\mu(\pi/4)} \bar{H} \ket{\Omega_\mu(\pi/4)}
- \frac{1}{2}E_\mu
- \frac{1}{2}E_0,
\label{eq:res_measure}
\end{equation}
where $E_0 = \bra{\Phi_0} \bar{H} \ket{\Phi_0}$ and $E_\mu = \bra{\Phi_\mu} \bar{H} \ket{\Phi_\mu}$.
All of these quantities are expectation values of $\bar{H}$ with respect to reference states that are easily generated with short quantum circuits.
The evaluation of the exact residual via Eq.~\eqref{eq:res_measure} has a cost similar to the evaluation of exact gradients in VQE via the shift rule \cite{schuld2019evaluating, kottmann2020feasible}.

\subsection{Efficient operator selection}
\label{sec:selection}

In this section, we generalize the PQE method to utilize a flexible dUCC ansatz built iteratively using a full operator pool.
As shown in the case of VQE \cite{grimsley2019adaptive, ryabinkin2020iterative}, significantly more compact and flexible approximations may be achieved if the operators that define the unitary are selected according to an importance criterion.
To formulate a selected version of the PQE approach, we propose to combine information about the residual with a cumulative importance criterion.
Since the residuals $r_\mu$ are zero for an eigenstate, we propose to estimate the importance of the operators $\hat{\kappa}_\mu$ using the magnitude of the residual ($|r_\mu|$).
However, instead of evaluating the importance of all the operators in the pool via operator averaging (like in gradient-based selection schemes \cite{grimsley2019adaptive, ryabinkin2020iterative}), we propose to sample a quantum state whose probability amplitudes encode the importance of \textit{all} operators up to rank $N$.

Suppose that we have determined a unitary $\hat{U}$ that satisfies the  residual condition $r_\mu = 0$ for all $\hat{\kappa}_\mu$ in the current ordered set $\mathcal{A}$.
In our approach we prepare a (normalized) quantum state of the form $
\ket{\tilde{r}}  = \tilde{r}_0 \ket{\Phi_0} + \sum_\mu \tilde{r}_\mu \ket{\Phi_\mu}$,
where the quantities $ \tilde{r}_\mu $ are approximately proportional to the residuals $r_\mu$.
When $\ket{\tilde{r}}$ is represented in a qubit basis, there is a one-to-one mapping between elements of the computational basis and the states $\Phi_\mu$.
Therefore, a measurement of the state $\ket{\tilde{r}}$ in the
qubit basis will yield one of the states $\Phi_\mu$ with probability $P_\mu = |\tilde{r}_\mu|^2$.
Repeated measurement of the state $\ket{\tilde{r}}$ provides a way to approximately determine the elements of the residual $r_\mu$ with the largest magnitude, and the corresponding operators $\hat{\kappa}_\mu$ that should be included in the unitary.
 When this strategy is combined with an efficient way to prepare the state $\ket{\tilde{r}}$, it is much more cost effective than evaluating all the elements of $r_\mu$ not included in $\mathcal{A}$ via operator averaging [Eq.~\eqref{eq:res_measure}].

Construction of the state $\ket{\tilde{r}}$ would require one to apply the Hamiltonian, which is not a unitary operator.
Therefore, we evaluate $\tilde{r}$ using the unitary operator $e^{-i \Delta t \hat{H}} = 1 -i \Delta t \hat{H} + \mathcal{O}(\Delta t^2)$ instead of $\hat{H}$.
By choosing a small time step, we can ensure that the nonlinear terms and errors due to the approximate implementation of $e^{-i \Delta t \hat{H}}$ (e.g., via Trotterization) do not affect the residual to leading order in $\Delta t$.
The residual state can then be defined as
\begin{equation}
    \begin{split}
        \ket{\tilde{r}} &= \Ud e^{i \Delta t \hat{H}} \hat{U} \ket{\Phi_0} \\
        &=  (1 + i\Delta t \Ud \hat{H} \hat{U})  \ket{\Phi_0} + \mathcal{O}(\Delta t^2).
    \end{split}
\end{equation}
The time-evolution operator may be approximated via Trotterization \cite{trotter1959product, suzuki1993improved} in combination with low-rank representations of the Hamiltonian \cite{berry2019qubitization}.
With a sufficiently large number of measurements $M$ of the state $\ket{\tilde{r}}$, we may approximate the values of the (normalized) squared residuals as
\begin{equation}
\label{eq:approx_res_sq}
\norm{\tilde{r}_\mu} \approx \frac{N_\mu }{ M },
\end{equation}
where $N_\mu$ is the number of times the state $\ket{\Phi_\mu}$ is measured.
Encoding the residual in a single quantum state allows us to efficiently sample the \textit{entire} operator pool without the need to generate and store individual elements of the residual vector in memory.
This distinctive feature makes it possible to employ this selection procedure with an operator pool that includes particle-hole excitation/de-excitation operators of arbitrary order.

To select important operators, we adopt a cumulative threshold approach that allows us to add a batch of operators at a time.
Our goal is to iteratively construct a unitary that contains the fewest operators.
This is realized with a selection procedure that adds the operators with the largest value of $\tilde{r}_\mu$ to $\mathcal{A}$ (as motivated by the Gershgorin circle theorem bounds discussed in Sec.~\ref{sec:pqe_theory}), and excludes all other operators in such a way that sum of their residuals squared is less than a threshold $\Omega^2$.
Specifically, we enforce that
\begin{equation}
\sum_{\hat{\kappa}_\mu \notin \mathcal{A}}^\text{excluded} \norm{r_\mu} \approx \sum_{\hat{\kappa}_\mu \notin \mathcal{A}}^\text{excluded} \frac{\norm{\tilde{r}_\mu}}{\Delta t^2} \leq \Omega^2,
\label{eq:op_thresh}
\end{equation}
where we have used the fact that $|\tilde{r}_\mu| \approx \Delta t | r_\mu|$.
In practice, we sort the operators in ascending order according to $\norm{\tilde{r}_\mu}$, and starting from the first element, we discard operators until Eq.~\eqref{eq:op_thresh} is satisfied. The remaining operators are appended to the end of $\mathcal{A}$ in order of decreasing $\norm{\tilde{r}_\mu}$.
The resulting operator ordering is consistent with the following renormalization transformation of the Hamiltonian
\begin{equation}
\begin{split}
\hat{H} & \rightarrow
 e^{-t_{\mu_{1}} \hat{\kappa}_{\mu_{1}}} \hat{H} e^{t_{\mu_1}} = \bar{H}_1 \\
& \rightarrow
 e^{-t_{\mu_{2}} \hat{\kappa}_{\mu_{2}}} \bar{H}_1 e^{t_{\mu_{2}} \hat{\kappa}_{\mu_{2}}} = \bar{H}_2 \\
& \rightarrow\cdots
\end{split}
\end{equation}
which begins with the largest many-body rotation and gradually continues with smaller ones.
Note that this ordering is the reverse of the one used in ADAPT-VQE and iQCC, where operators with the largest gradient are applied first to the reference.
In applications of this selection scheme, we found that our operator ordering is more numerically robust and accurate than its reverse.
This selection procedure is easily integrated in the PQE algorithm by performing a series of computations with increasingly larger ordered sets.
When no new operators are added to $\mathcal{A}$, the computation is considered converged, and the final operator set satisfies Eq.~\eqref{eq:op_thresh}.
The details of the selected PQE algorithm are discussed in Sec.~\ref{sec:algorithm}.

Throughout this work, we ignore errors that arise from finite measurement of the approximate residual $\ket{\tilde{r}}$.
However, since operator selection only requires an approximate determination of the $|\tilde{r}_\mu|^2$ values, it is not necessary to perform a large number of measurements.
Indeed, in Appendix~\ref{apdx:aa_reduce_cost}, we discuss a simple strategy based on sampling $\ket{\tilde{r}}$ a fixed number of times that performs as well as the exact scheme.

\subsection{Outline of the selected PQE algorithm}
\label{sec:algorithm}

\begin{figure*}[ht!]
\centering
\includegraphics[width=5.5in]{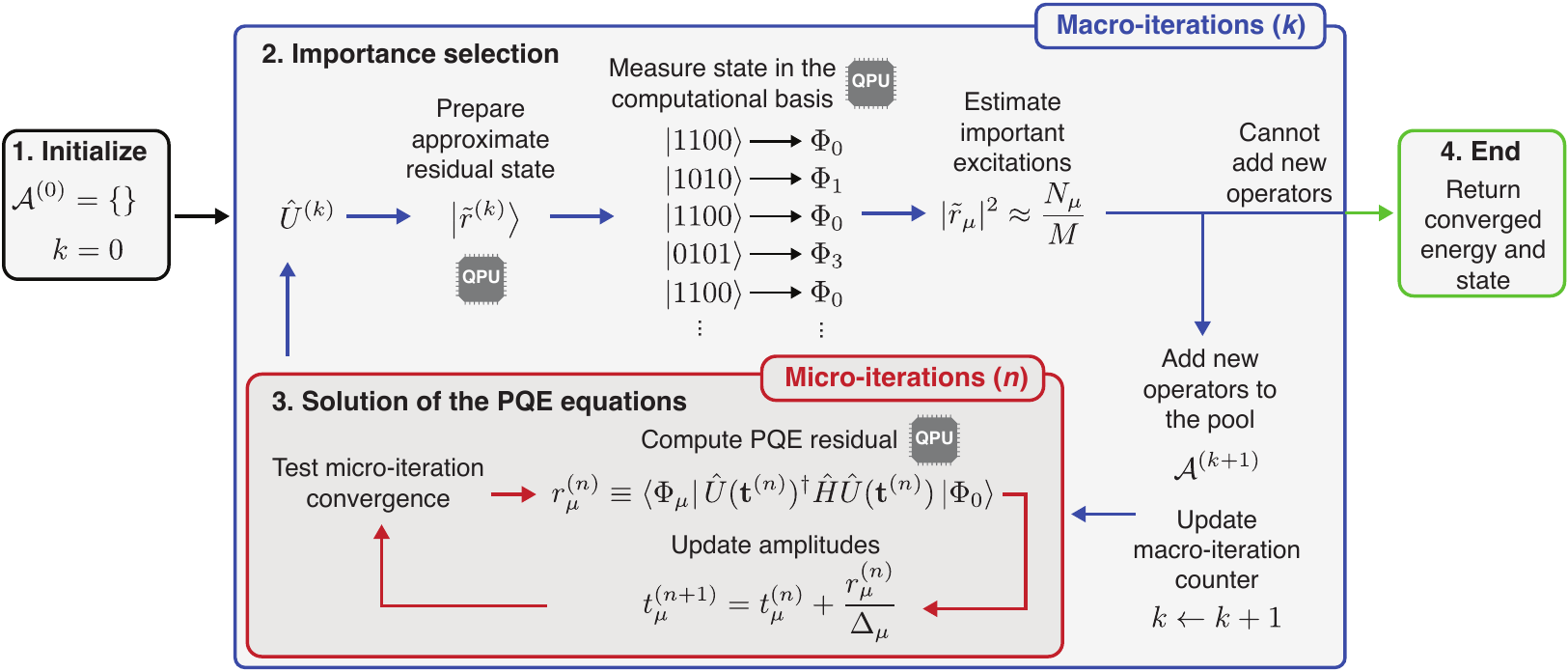}
\caption{Outline of the adaptive PQE algorithm. Steps labeled ``QPU'' indicate parts of the algorithm that run on a quantum processing unit.}
\label{fig:algorithm}
\end{figure*}

The combination of the PQE approach  with the selection procedure described in Sec.~\ref{sec:selection} leads to a very efficient  flexible ansatz quantum algorithm, which we refer to as selected PQE (SPQE).
The selected PQE algorithm requires the simultaneous solution of the residual conditions and the selection of important excitation operators.
To realize this scheme, we alternate micro-iterations to converge the residual equations (for the current ordered  operator set) with macro-iterations that perform importance selection of new operators.
The SPQE procedure is illustrated in Fig.~\ref{fig:algorithm} and consists of the following steps:
\begin{enumerate}
\item \textbf{Initialization}. The user provides the occupation numbers that define the reference state $\Phi_0$. Start at macro-iteration number $k = 0$ with an empty operator set ($\mathcal{A}^{(0)} = \{ \}$).

\item \textbf{Importance selection}. At macro-iteration $k$, perform $M$ measurements in the computational basis for the the state $\ket{\tilde{r}^{(k)}} = \Ukd e^{i \Delta t \hat{H}} \Uk \ket{\Phi_0}$.
The number of times the state $\Phi_\mu$ is measured is accumulated in the variable $N_\mu$. These numbers are used to estimate the square residuals via Eq.~\eqref{eq:approx_res_sq}, which are in turn used to select important excitations not included in the current pool $\mathcal{A}_{k}$.
The current operator set $\mathcal{A}_{k}$ and all the new selected operators are included in the new set $\mathcal{A}_{k + 1}$.
When forming this new ordered set, we append the new operators---sorted in decreasing value of the approximate squared residual---to $\mathcal{A}_{k}$.
If the sum in Eq.~\eqref{eq:op_thresh} over \textit{all} approximate residuals is less than the threshold $\Omega^2$, such that no new operators are added, then return the final energy.

\item \textbf{Solution of the PQE equations}. Using the new set $\mathcal{A}_{k + 1}$, solve the PQE equations via quasi-Newton micro-iterations.
These micro-iterations alternate the evaluation of the residuals [Eq.~\eqref{eq:ucc2}] and the amplitude update [Eq.~\eqref{eq:fixed_point}].
The micro-iterations are considered converged when the norm of the residual vector $\normnorm{\mathbf{r}}$ is less than a user specified threshold $\omega_r$.
Once converged, this step produces a new set of amplitudes [$\mathbf{t}^{(k + 1)}$] and the corresponding unitary [$\hat{U}^{(k + 1)}$], as well as the updated energy [$E^{(k + 1)}$].
Increase the macro-iteration number by one ($k \leftarrow k + 1$) and go to Step 2.
\end{enumerate}

All VQE and PQE methodologies were implemented in a development branch of the open source package \textsc{QForte}
, and utilize its state--vector simulator.
All VQE calculations use a micro-iteration convergence threshold $\omega_g = 10^{-5}$~\Eh for the gradient norm $\normnorm{\mathbf{g}}$ and all PQE calculations use a micro-iteration threshold $\omega_r = 10^{-5}$~\Eh for the residual norm $\normnorm{ \mathbf{r} }$.
Note that when the $\hat{\kappa}_\mu$ operators are mapped to a qubit basis they can be expressed as a sum of Pauli operator strings that commute \cite{Romero:2019hk}, and therefore, a single operator $e^{t_\mu \hat{\kappa}_\mu}$ can be implemented exactly as a product of exponentials without invoking the Trotter approximation.

\section{Results and Discussion}
\label{sec:results}

\subsection{Comparison of PQE and VQE with a  disentangled UCC ansatz}

\begin{figure*}[ht!]
\centering
\includegraphics[width=6.0in]{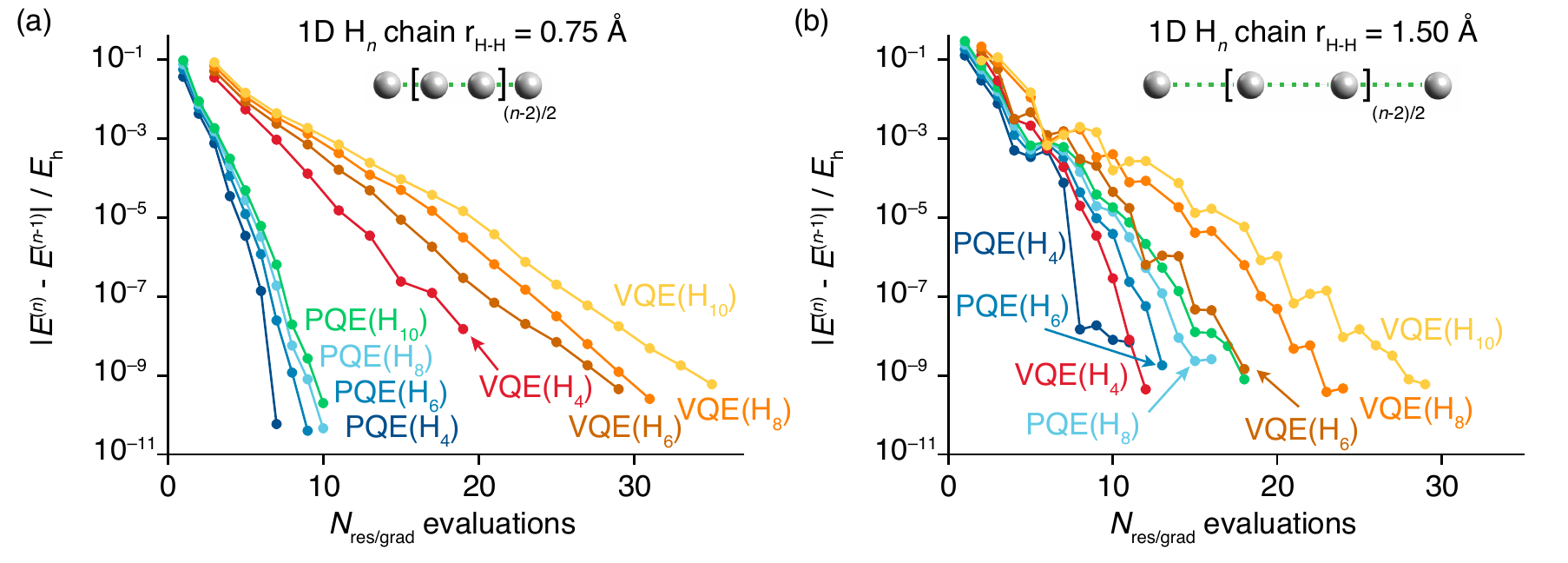}
\caption{dUCCSD  energy convergence for linear \ce{H4}--\ce{H10} chains in a STO-6G basis at (a) $r_{\rm{H-H}} = 0.75$~{\AA}, and (b) $r_{\rm{H-H}} = 1.50$~{\AA}. $| E^{(n)} - E^{(n-1)}|$  is the absolute value of the energy change between subsequent iterations. Both plots compare PQE~vs.~VQE convergence with respect to number of residual (for PQE) or gradient (for VQE) evaluations ($N_{\rm{res/grad}}$).}
\label{fig:Hn_econv_both}
\end{figure*}

Our initial goal is to compare the performance of PQE and VQE using a unitary coupled-cluster trial state truncated at a given particle-hole excitation level.
We test these two methods on a family of linear hydrogen chains (with identical nearest-neighbor distance) ranging from four to ten atoms, both near their equilibrium geometries ($r_\text{H-H}  = 0.75$ \AA{}) and stretched geometries ($r_\text{H-H}  = 1.5$ \AA{}).
Hydrogen models such as these have been studied experimentally with VQE \cite{google2020hartree} and have recently been used as a benchmark for both quantum \cite{grimsley2019adaptive, Stair_2020} and classical \cite{Motta2017TowardsThe, motta2020ground, stair2020exploring} algorithms.

Figure~\ref{fig:Hn_econv_both} shows the energy convergence of PQE and VQE using a disentangled UCC ansatz with singles and doubles (dUCCSD).
All calculations employed restricted Hartree-Fock (RHF) orbitals from the quantum chemistry package \textsc{Psi4} \cite{smith2020psi4}, and Pauli-operator Hamiltonians obtained via the Jordan--Wigner transformation implemented in \textsc{QForte} \cite{stairqforte}.
To achieve optimal performance for both VQE and PQE, we employ the Broyden--Fletcher--Goldfarb--Shannon (BFGS) algorithm \cite{broyden1970convergence, fletcher1970new, goldfarb1970family, shanno1970conditioning}  (as implemented in the \textsc{SciPy} \cite{virtanen2020scipy} scientific computing library) with analytical gradients for VQE, and  DIIS \cite{pulay1980convergence,SCUSERIA1986236} to accelerate amplitude convergence of PQE.
These computations use the same operator ordering for both approaches, with all amplitudes initialized to zero.
The ordering of the operators $e^{t_{\mu_i} \hat{\kappa}_{\mu_i}}$ entering Eq.~\eqref{eq:qucc} is defined by the binary representation of the corresponding determinants $\ket{\Phi_{\mu_i}} = \hat{\tau}_{\mu_i} \ket{\Phi_0}$ in the occupation number representation.
Because the  disentangled UCCSD state cannot exactly parameterize an eigenstate of the Hamiltonian, the numerically converged PQE and VQE energies are not identical.
Nevertheless, for all the cases we examined the converged PQE and VQE energies differ by less than $10^{-6}$~\Eh.

Near the equilibrium geometry, we find that the PQE energy converges significantly faster than the VQE energy with number of residual vs. gradient evaluations, respectively.
For example, to converge the near-equilibrium \ce{H10} energy to $10^{-6}$~\Eh, PQE requires only seven residual evaluations, while VQE requires approximately 23 gradient evaluations.
In the case of VQE, we also observe that the number of required gradient evaluations grows with system size, with \ce{H10} taking twice as many gradient vector evaluations than \ce{H4} to converge.
On the contrary, PQE computations converge with similar speed for all equilibrium hydrogen systems.
Plots of the energy change vs. the norm of the residual/gradient vector show similar trends and are reported in Appendix~\ref{sec:add_num_vqe_compare}.

At the stretched geometry, strong correlation effects cause the  disentangled UCCSD trial state to perform more poorly, with both PQE and VQE, yielding energy errors that range from 1.39~m\Eh (for \ce{H4}) to 13.59~m\Eh (for \ce{H10}).
We see that PQE converges slightly more slowly in the stronger correlation regime, with stretched \ce{H10} requiring 13 residual evaluations (instead of seven) to converge the energy to $10^{-6}$~\Eh.
However, PQE always converges faster than VQE, with the latter requiring 19 gradient-vector evaluations to converge stretched \ce{H10} to the same accuracy level.
We also find similar trends in PQE and VQE convergence for \ce{BeH2}, for which convergence data can be found in Appendix~\ref{sec:add_num_vqe_compare}.

In summary, this initial set of results suggests that for a given trial state, optimization via PQE is faster than VQE and less dependent on the number of parameters to optimize.
We expect this to be the case also for VQE based on numerical gradients or gradient-free optimization methods, since these two variants are known to be slower compared to the the BFGS approach adopted here \cite{Romero:2019hk}.

\subsection{Effect of stochastic errors on the convergence of PQE and VQE}

\begin{figure*}[ht!]
\centering
\includegraphics[width=3.375in]{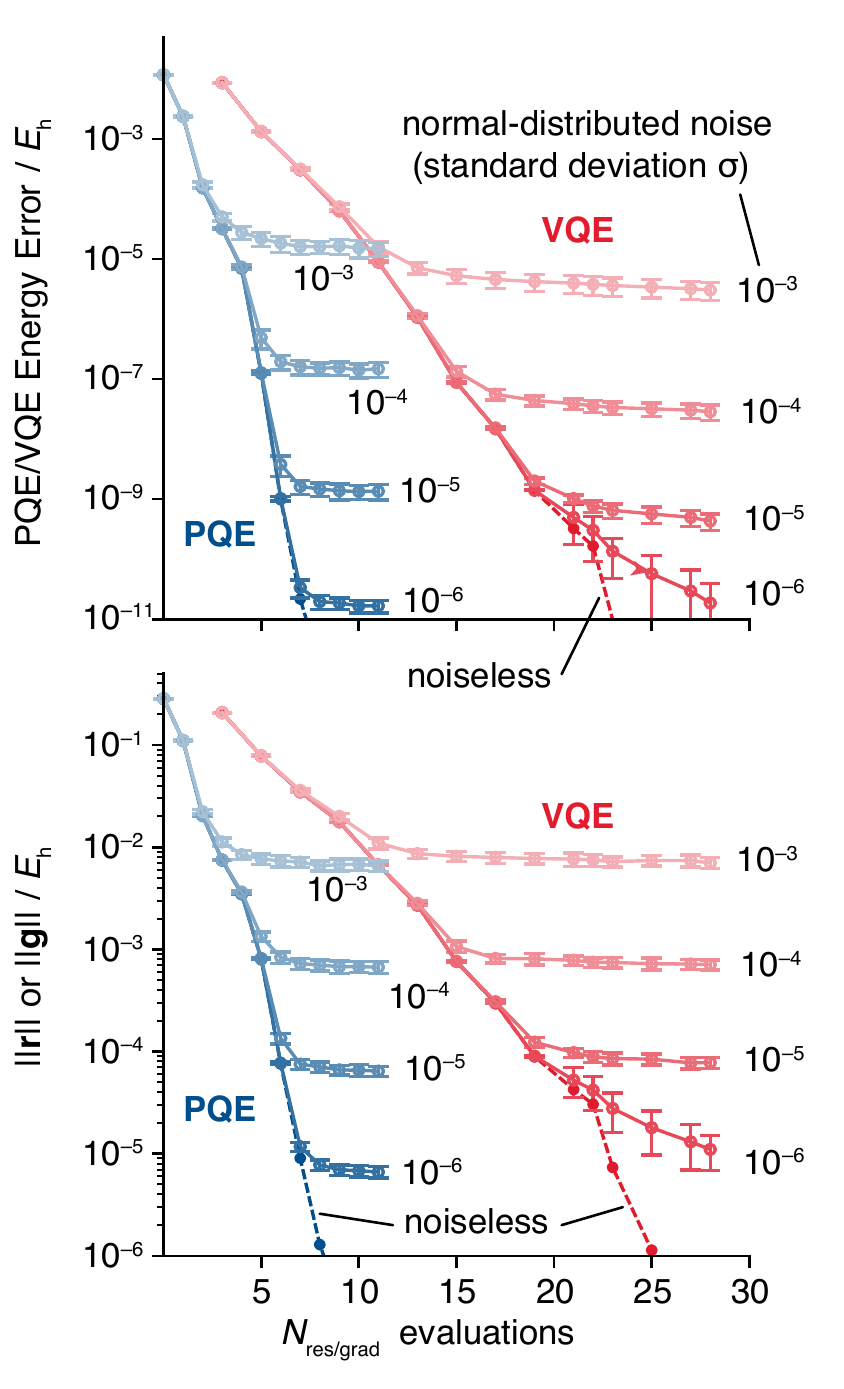}
\caption{Energy and residual/gradient norm ($\normnorm{\mathbf{r}}$)/($\normnorm{\mathbf{g}}$) convergence of dUCCSDTQ  wave functions optimized with PQE/VQE with various amounts of stochastic noise added to the residuals/gradients. Energy error is relative to FCI.
$\sigma$ controls the degree of noise and is the standard deviation of the normal distribution, centered at the exact residual/gradient value, from which all residuals/gradients used in the calculations are randomly sampled [see Eq.~\eqref{eq:res_noise}]. Values at each PQE/VQE iteration are averages over 50 runs on \ce{H4} at $r_{\mathrm{H-H}} = 1.0$~\AA. Error bars denote one standard deviation.}
\label{fig:H4_w_noise}
\end{figure*}

The results presented in the previous section assumed error-free quantum gates and arbitrarily precise measurements.
In practice, calculations performed on NISQ hardware are affected by decoherence errors, poor gate fidelity, readout errors, and loss of precision due to insufficient measurements.
These sources of error will lead to incorrect gradients and residuals that are then passed to a classical optimizer.
Therefore, it is interesting to compare the resilience of PQE and VQE procedures when the residuals and gradients are affected by stochastic errors.

To model the presence of errors, we modify the PQE procedure by adding to the residual vector a stochastic error sampled from a Gaussian distribution with standard deviation $\sigma$ [$\mathcal{N}(0,\sigma^2)$]
\begin{equation}
\label{eq:res_noise}
r_\mu^\text{measured} = r_\mu + \mathcal{N}(0,\sigma^2).
\end{equation}
For VQE, we similarly add stochastic noise to the exact energy gradients.
Using Eq.~\eqref{eq:res_noise} as a noise model mainly emulates errors that arise from finite measurement.
Because the inexact residuals $r_\mu^\text{measured}$ are used to update the cluster amplitudes via Eq.~\eqref{eq:fixed_point}, this noise model also gives rise to control errors, or errors that refer to the difference between the unitiaries for noiseless updated amplitudes $\hat{U}(\mathbf{t})$ and noisy updated amplitudes $\hat{U}(\mathbf{t}+\Delta\mathbf{ t})$.
Control errors due to finite measurement have been modeled this way in previous studies \cite{Romero:2019hk} and will always propagate through optimization on physical hardware.
We note, however, that using Eq.~\eqref{eq:res_noise} exclusively as a noise model is insufficient to capture more nuanced or device-specific errors such as decoherence.
We tested the performance of PQE and VQE under noise by performing a batch of 50 computations on the linear \ce{H4} molecule with nearest-neighbor distance set to 1.0~{\AA}.
We use the  disentangled UCC ansatz with up to quadruple excitations, which spans the full operator set for this system.

Figure~\ref{fig:H4_w_noise} shows a comparison of PQE and VQE optimized with various levels of noise, controlled by the magnitude of $\sigma$ in Eq.~\eqref{eq:res_noise}.
We find that for all values of $\sigma > 0$, the energy convergence of PQE is essentially identical to that of noiseless PQE until some point, after which the energy error hovers around a finite value.
Similar behavior can be seen for convergence of the residual vector.
We find that VQE has similar characteristics to PQE in the presence of noise, but that it is able to achieve slightly more accurate energies for a given $\sigma$ value.
With the same noise level, however, VQE generally requires two to three times the number of gradient evaluations as the number of residual evaluations required by PQE.
An important aspect of this comparison is that, for a given $\sigma$, both the residual and gradient vectors yield comparable asymptotic errors in PQE and VQE, respectively.
Conversely, PQE and VQE computations of the comparable energy accuracy require similar precision in the measurement of the residual and gradient vectors.
In Appendix~\ref{sec:formal_vqe_compare} we use this result to estimate the relative cost of PQE and VQE via a formal analysis.

\subsection{Selected PQE based on a full dUCC operator pool}
\begin{figure*}[ht!]
\centering
\includegraphics[width=6.5in]{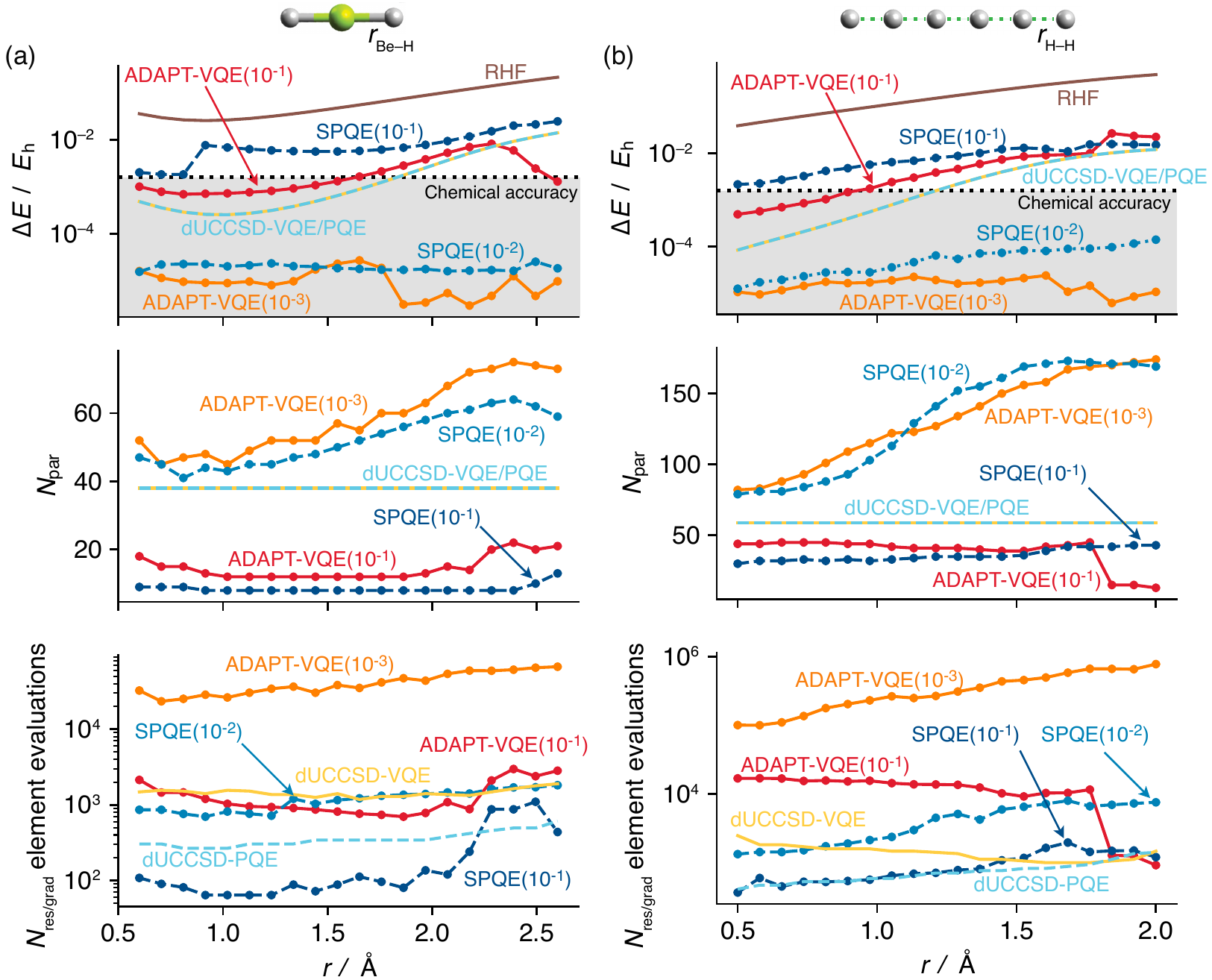}
\caption{Ground state potential energy curve for the symmetric dissociation of (a) \ce{BeH2} and (b) \ce{H6} computed using a minimal (STO-6G) basis. The energy error relative to FCI (top), number of classical parameters used (middle), and number of individual elements of the gradient (for VQE) or residual (for PQE) evaluated (bottom) are given as a function of the \ce{Be}--\ce{H} and \ce{H}--\ce{H} bond length. Here, ADAPT-VQE uses a generalized singles and doubles operator pool and is optimized with the BFGS algorithm, and gradient convergence thresholds $10^{-1}$ ($\epsilon_1$) and $10^{-3}$ ($\epsilon_3$). SPQE results use macro-iteration convergence thresholds $\Omega=10^{-1}$ and $10^{-2}$. The top plots also show the energy error corresponding to chemical accuracy, here defined as 1 kcal/mol $\approx$ 1.59 m\Eh.}
\label{fig:H6_BeH2_pes_compare}
\end{figure*}

Here we compare the results of selected PQE (SPQE) with an arbitrary order particle-hole operator pool, and ADAPT-VQE, which unless otherwise noted, uses a generalized singles and doubles pool (operators of the form $\cop{p}\aop{q}$ and $\cop{p}\cop{q}\aop{s}\aop{r}$, where the indices $p,q,r,s$ run over all spin orbitals).
To test these methods, we compute the energy as a function of bond distances for: 1) the symmetric dissociation of the linear \ce{BeH2} molecule and 2) the symmetric dissociation of a chain of six hydrogen atoms.
In both cases, there is a build up of strong correlation effects as the bond length increases.
For each system, we report two sets of results for SPQE using the cumulative threshold $\Omega = 10^{-1}$ and $10^{-2}$~\Eh, and two sets of ADAPT-VQE results with gradient threshold $10^{-1}$ and $10^{-3}$~\Eh ($\epsilon_1$ and $\epsilon_3$ in the original notation used by the authors).
Since we later compare these methods to classical approaches formulated in a determinant basis, we do not employ a pool of spin-adapted operators.

The dissociation curve of \ce{BeH2} shown in Fig.~\ref{fig:H6_BeH2_pes_compare} (a) demonstrates that both SPQE and ADAPT-VQE are able to achieve significantly smaller energy errors than dUCCSD, while using only 10--20 more parameters.
Although the two approaches employ different selection schemes, the ADAPT-VQE($\epsilon_3$) and SPQE($\Omega = 10^{-2}$) approaches produce compact trial states with a similar number of classical parameters and comparable errors.
The same trends are seen in symmetric dissociation curve of \ce{H6}, which is shown in Fig.~\ref{fig:H6_BeH2_pes_compare} (b).
However, for this system we find that achieving sub-m\Eh accuracy---particularly with the onset of strongly correlation at $r_{\rm{H-H}}$ values greater than 1.5 \AA{}, requires a number of parameters that approaches the size of the full Hilbert space (200) for both ADAPT-VQE and SPQE.
The need to saturate Hilbert space to accurately describe \ce{H6} is likely due to how small this example is, and it speaks more to the ability of the trial states to produced compact representations rather than the performance of these algorithms in optimizing such ans\"{a}tze.

The most noticeable difference between SPQE and ADAPT-VQE can be seen in the bottom panel of Fig.~\ref{fig:H6_BeH2_pes_compare} (a)-(b): for trial states with comparable number of parameters, SPQE requires significantly fewer residual element evaluations than gradient element evaluations in ADAPT-VQE.
For example, at a \ce{Be}--\ce{H} bond distance of 1.65~\AA{}, both methods produce very similar energy errors using almost the same number of parameters, but ADAPT-VQE($\epsilon_3$) requires the evaluation of 35155 elements of the gradient, whereas SPQE($\Omega = 10^{-2}$) requires only 1220 elements of the residual.
Importantly, we note that the bottom panels of Fig.~\ref{fig:H6_BeH2_pes_compare} exclusively count the number of elements of the gradient or residual required by the optimization, and do not include the additional measurements required for operator selection.

\begin{table*}[!ht]
\centering
\caption{Ground state of \ce{H6} computed using a minimal (STO-6G) basis with RHF orbital convergence threshold of $10^{-10}$~\Eh. Comparison of SPQE with threshold $\Omega$ and ADAPT-VQE using the same number of parameters as SPQE. ADAPT-VQE results are computed for both a generalized singles and doubles operator pool (GSD) and a particle hole singles and doubles pool (SD). The properties reported are the energy error with respect to FCI [$\Delta E$, in \Eh], the number of classical parameters used [$N_{\rm{par}}$], the number of parameters corresponding to three-body or higher excitations [$N_{\rm{T+}}$], the number of CNOT gates used in the unitary [$N_{\rm{CNOT}}$] (not optimized), and the total number of residual or gradient element evaluations [$N_\mathrm{res}$ or $N_\mathrm{grad}$]. $r$ denotes the H-H nearest neighbor distance in \AA{}ngstrom.}

\scriptsize 

\begin{tabular*}{\textwidth } 
{@{\extracolsep{\stretch{1.0}}}*{1}{c}*{13}{r}@{}}
    \hline

    \hline
& \multicolumn{5}{c}{SPQE ($\Omega = 10^{-1}$ \Eh) } & \multicolumn{4}{c}{ADAPT-VQE-GSD } & \multicolumn{4}{c}{ADAPT-VQE-SD } \\
    \cline{2-6}      \cline{7-10}  \cline{11-14}
     $r$  & $\Delta E$  & $N_{\rm{par}}$  & $N_{\rm{T+}}$ & $N_{\rm{CNOT}}$ &  $N_\mathrm{res}$  & $\Delta E$ & $N_{\rm{par}}$ &  $N_{\rm{CNOT}}$ &  $N_\mathrm{grad}$  & $\Delta E$ & $N_{\rm{par}}$ &  $N_{\rm{CNOT}}$ & $N_\mathrm{grad}$ \\
    \hline

  0.50 &  0.002153  &  30   &  0 &   2400 & 339 &   0.002152 & 30  & 2400          &  8378    &   0.002152 & 30 & 2400           &  8378 \\
  1.00 &  0.006050  &  32   &  0 &   2720 & 503 &   0.005872 & 32  & 2720          &  8399    &   0.005872 & 32 & 2720 &  8399 \\
  1.50 &  0.012487  &  36   &  0 &   2944 & 1103 & 0.011176  & 36  & 3040          & 8046     &    0.011176 & 36 & 3040          &  8046 \\
  2.00 &  0.015066  &  43   &  8 &         20272  & 1087 & 0.011204  & 43 & 3560 & 15176   &  0.010350 & 43 & 3312           &  12592 \\[6pt]

& \multicolumn{5}{c}{SPQE ($\Omega = 10^{-2}$ \Eh) } & \multicolumn{4}{c}{ADAPT-VQE-GSD } & \multicolumn{4}{c}{ADAPT-VQE-SD } \\
    \cline{2-6}      \cline{7-10}  \cline{11-14}
     $r$  & $\Delta E$  & $N_{\rm{par}}$  & $N_{\rm{T+}}$ & $N_{\rm{CNOT}}$ &  $N_\mathrm{res}$  & $\Delta E$ & $N_{\rm{par}}$ &  $N_{\rm{CNOT}}$ & $N_\mathrm{grad}$ & $\Delta E$ & $N_{\rm{par}}$ &  $N_{\rm{CNOT}}$ & $N_\mathrm{grad}$  \\
    \hline

  0.50 &          0.000013    &   79   &    24    &          15568  &           1127     &         0.000012     &   79 &   6608 &   87506     &          0.000085     &   79  &          4608     &             29707 \\
  1.00 &          0.000031    & 105   &    46    &          33232  &  2076    & 0.000033   & 105 &   9244 &   166119    & 0.000064    & 105  &  8128    &  388182 \\
  1.50 &  0.000079  & 166   &  111    & 166032 & 6074    & 0.000004   & 166 & 14528 &   566143    & 0.000028    & 166  & 12576   &  1437917 \\
  2.00 &          0.000141   &  169  &  114    &          226768 & 9537    & 0.000018   & 169 & 14776 &           719390    &           0.000096     & 169  & 12944   &          1312127 \\[3pt]

    \hline

    \hline
\end{tabular*}
\label{tab:N_CNOT_comparison_2}
\end{table*}

Since ADAPT-VQE and SPQE select new operators from their pools using different importance criteria, it is not possible to perform a direct comparison of their performance using fixed thresholds.
To facilitate this comparison, in Table \ref{tab:N_CNOT_comparison_2} we report SPQE results using two values of $\Omega$ ($10^{-1}$ and $10^{-2}$~\Eh) together with ADAPT-VQE results obtained using an ansatz with the same number of parameters as SPQE.
We include ADAPT-VQE results obtained using both a generalized singles and doubles operator pool (GSD), and a particle-hole singles and doubles pool (SD).
The results in Tab.~\ref{tab:N_CNOT_comparison_2} obtained with $\Omega$ = $10^{-1}$~\Eh show SPQE and the two variants of ADAPT-VQE to perform equally well at all bond distances.
The second set of results obtained with a tighter threshold ($\Omega=10^{-2}$) show similar performance of the methods at short bond lengths, with two notable exceptions.
First, at $r_{\rm{H-H}} = 2.0$~\AA{} ADAPT-VQE-GSD yields more accurate results than ADAPT-VQE-SD and SPQE.
At this point, the SPQE ansatz contains 55 singles and doubles, and 114 operators of higher rank, yielding an error of about 0.14~m\Eh, while ADAPT-VQE-GSD is  an order of magnitudes more accurate.
Second, also at $r_{\rm{H-H}} = 2.0$~\AA{}, SPQE uses only  9537 residual element evaluations, while ADAPT-VQE requires 719390 (GSD) and 1312127 (SD) gradient element evaluations.
The evaluation of fewer residual elements in SQPE will correspond to approximately the same savings in total number of measurements (see Appendix~\ref{sec:formal_vqe_compare}).

A final important aspect to compare between SPQE and ADAPT-VQE is the number of native CNOT gates, which we consider as a proxy for circuit depth.
Tab.~\ref{tab:N_CNOT_comparison_2} reports the number of CNOT gates for the converged trial states.
These numbers overestimate the actual gate count since they ignore optimizations such as the cancellation of Jordan--Wigner strings \cite{Haystings2015Improving}, especially for three- and higher-body operators.
We see that at the larger threshold values ($\Omega=10^{-1}$ for SPQE and $\epsilon_1=10^{-1}$ for ADAPT) the number of CNOTs for all three approaches are relatively similar, and are generally within a factor of 2 of one another.
However, at tighter thresholds the SPQE circuit contains significantly more CNOT gates than the one for ADAPT-VQE.
For example, $r_{\rm{H-H}} = 2.0$~{\AA}), 114 of the 169 operators used in SPQE are three-body or higher, while ADAPT-SD and ADAPT-GSD of course only contain only up to two-body operators.
Consequently, the SPQE($\Omega=10^{-2}$) circuit contains more CNOT gates (226768) than ADAPT-VQE-GSD  (14776).
As discussed in Appendix~\ref{sec:formal_vqe_compare}, this large difference in CNOT count is due to the growth in cost to implement the exponential of $n$-body second-quantized operators as a function of $n$ (and the lack of circuit optimization).
Nevertheless, it is important to note that the systems studied here are not large enough to draw definitive conclusions about the relative performance of SPQE and ADAPT-VQE.
For example, the pool of generalized singles and doubles (GSD) for \ce{H6} contains 870 operators, a number significantly larger than the size of the full Hilbert space (200).
This scenario is unlike most systems of interest, where the number of generalized singles and doubles is much less than the number of particle-hole operators.
Unfortunately, our attempts to compare numbers for systems larger than \ce{BeH2} and \ce{H6} were unsuccessful due to the high computational cost of simulating ADAPT-VQE.

\begin{figure}[h!]
\centering
\includegraphics[width=3.35in]{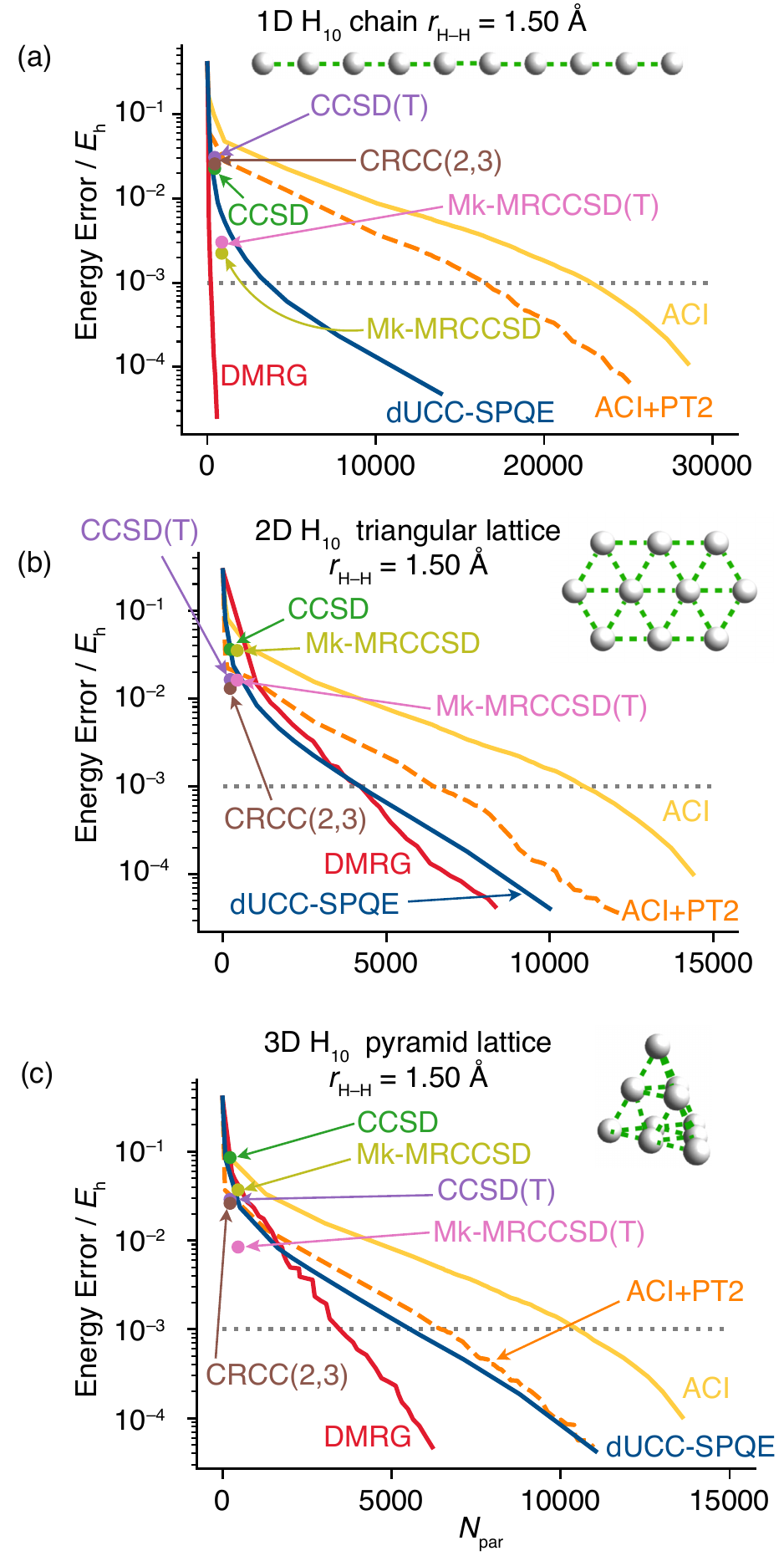}
\caption{Results for the 1D, 2D, and 3D \ce{H10} models use a STO-6G basis at $r_{\rm{H-H}}=1.50$~{\AA}. Singlet ground state energy errors relative to FCI as a function of number of variational parameters $N_{\rm{par}}$ for dUCC-SPQE, ACI, and DMRG. Unsigned energy errors for methods with fixed number of parameters (taken to be the number of cluster amplitudes) are shown by colored dots. Energy errors ACI with a second order perturbative correction (ACI+PT2) are also shown by orange dotted lines. For ACI+PT2, CCSD(T), CRCC(2,3), and Mk-MRCCSD(T) we only count the number of nonperturbative amplitudes. The accuracy volume threshold [0.1~m\Eh per electron] is plotted as a grey dotted line.}
\label{fig:H10_123D_econv}
\end{figure}

A second aspect we investigate, is the ability of the selected PQE approach to compactly represent wave functions for systems displaying strong correlation effects, were many-body methods commonly fail due to the breakdown of the mean-field approximation.
We compare the performance of SPQE with the adaptive configuration interaction \cite{Schriber2016Adaptive} (ACI) and DMRG using data generated in our recent benchmark study of hydrogen systems \cite{stair2020exploring}.
In this work, we characterize the resource requirements of a computational method $X$ with the \textit{accuracy volume} ($\mathcal{V}_X$), defined as the smallest number of parameters necessary to achieve a given energy error per electron (here taken to be $10^{-4}$ \Eh/electron).
We consider three \ce{H10} model systems: a one dimensional (1D) linear chain, a two dimensional (2D) triangular lattice and a three dimensional (3D) close-packed pyramid.
Using a minimal STO-6G basis and a 1 qubit to 1 spin-orbital mapping, the 1D, 2D, and 3D \ce{H10} models are represented with $2^{20}$ computational basis states, but are restricted to 31752 and 15912, and 15912 determinants, respectively, after accounting for spin and Abelian point-group symmetries.

Figure~\ref{fig:H10_123D_econv} (a) displays the SPQE, ACI, and DMRG energy error as a function of classical variational parameters for the 1D \ce{H10} system at a stretched bond length of $r_{\rm{H-H}}=1.50$~{\AA}.
DMRG, which is ideally suited to simulate gapped 1D systems, affords the most compact wave function for the 1D chain, with an accuracy volume of only 176 variational parameters ($\mathcal{V}_{\rm{DMRG}}^{\rm{1D}} = 176$).
The SPQE exponential ansatz is less compact than the DMRG matrix product state, with an accuracy volume $\mathcal{V}_{\rm{SPQE}}^{\rm{1D}} = 3510$ parameters.
We observe that the ACI wave function---a linear ansatz built by selecting the determinants with the largest energy contributions---gives the least compact representation, such that $\mathcal{V}_{\rm{ACI}}^{\rm{1D}} = 22989$ at the same level of accuracy.
When the ACI energy is augmented with a second-order perturbative correction---which accounts for determinants excluded from the  wave-function expansion---a more compact ansatz is sufficient to achieve an energy error of $10^{-3}$~\Eh.
This suggests that it might be valuable to formulate a classical perturbative correction to the SPQE energy to help capture even more correlation energy.

Results for the 2D \ce{H10} lattice [Fig.~\ref{fig:H10_123D_econv} (b)]  show that SPQE yields a more compact representation than DMRG or ACI until an energy error of approximately 1~m\Eh.
For the 2D system the accuracy volume follows the order SPQE (4193) $\approx$ DMRG (4218) $<$ ACI (11122).
At energy errors less than 1~m\Eh DMRG still affords the most compact representation for the 2D system.
Finally, for the 3D \ce{H10} pyramid lattice shown in Fig.~\ref{fig:H10_123D_econv} (c), we find that DMRG yields the most compact representation ($\mathcal{V}_{\rm{DMRG}}^{\rm{3D}} = 3549$), but not nearly by the same margin as for the 1D system.
Again, the SPQE has a significantly more compact wave function than ACI such that $\mathcal{V}_{\rm{SPQE}}^{\rm{3D}} = 5544$ and $\mathcal{V}_{\rm{ACI}}^{\rm{3D}} = 10 519$.
While further investigation of larger strongly-correlated systems will be necessary, it is encouraging to see that SPQE performs similarly to or better than two state-of-the-art classical methodologies when applied to the 2D and 3D \ce{H10} systems.

We have also included in Fig.~\ref{fig:H10_123D_econv} energy errors for several classical coupled-cluster variants.
Specifically we have included results from Ref.~\cite{stair2020exploring} for CC with singles and doubles excitations \cite{Purvis1982FullCoupled} (CCSD), CCSD with perturbative triples \cite{Raghavachari1989AFifth} [CCSD(T)], and the completely renormalized CC with perturbative triples \cite{piecuch2005renormalized} [CRCC(2,3)].
These CC methods are (generally) more computationally affordable than the other methods. In the 2D and 3D systems, CCSD performs comparably to dUCC-SQPE (with the same number of parameters), while CCSD(T) and CRCC(2,3) produce more accurate energies. However, all coupled-cluster energies are more than 10 m\Eh off from the FCI energy.
Fig.~\ref{fig:H10_123D_econv} also reports results computed using Mukherjee MRCC with singles and doubles \cite{Mahapatra:1999tm,Evangelista:2007hz} (Mk-MRCCSD)  and Mk-MRCCSD augmented with perturbative triples \cite{Evangelista:2010cq} [Mk-MRCCSD(T)] using an active space containing the highest occupied and lowest unoccupied Hartree--Fock orbitals.
The Mk-MRCC results improve upon single-reference coupled-cluster methods at the cost of doubling the number of cluster amplitudes. The Mk-MRCC methods are particularly accurate for the 1D system, where they produce errors of the order of 2--3 m\Eh.
Despite the improvement shown by the multireference CC methods, it is important to note that these methods have a computational cost that scales exponentially with the number of active space determinants.

\section{Conclusions}
\label{sec:conclusions}

In this work, we present a new NISQ-friendly algorithm---the projective quantum eigensolver (PQE)---to compute the ground state of a many-body problem using disentangled (factorized) unitary coupled-cluster trial states.
The PQE approach consists of a nonlinear optimization problem whose solution requires the evaluation of projections of the Schr\"{o}dinger equation onto a many-body basis (residual vector) but still gives energies that are a variational upper bound to the ground state energy.
We show how to efficiently evaluate the residual vector via measurement of simple expectation values, with a cost that is twice that of an energy evaluation (per element).
For small molecular systems, we find that PQE and VQE with a fixed dUCCSD  trial state converge to nearly identical energies; however, the number of residual evaluations required by PQE is smaller than that of the gradient evaluations needed by VQE.
PQE shows similar resiliency as VQE and still converges more rapidly in the presence of stochastic noise.

To treat strongly correlated electrons, we introduce a selected variant of PQE in which the trial state is constructed iteratively by adding batches of important operators.
The resulting SPQE algorithm can construct efficient unitary circuits like ADAPT-VQE, but it requires orders of magnitude fewer residual element evaluations than gradient element evaluations for the latter.
In SPQE, the selection of new operators is done according to the magnitude of the elements of the residual vector, and is performed by sampling a quantum state that directly encodes in its probability amplitudes the importance of the entire operator pool.
Because the selection cost in SQPE is not affected by the size of the operator pool, the unitary can include operators of rank up to the total number of particles.

Finally, we compare the energy convergence with the number of parameters for 1D, 2D, and 3D \ce{H10} lattices using SPQE and two classical methods well suited to treat strong electronic correlation: the adaptive configuration interaction  and the density matrix renormalization group.
Given a target accuracy of up to approximately 1~m\Eh, we find that SPQE produces significantly more compact trial states for the 2D system than ACI and is comparable to DMRG.
However, DMRG affords the most compact wave function parameterization in 1D, for accuracies below 1~m\Eh in 2D, and by a much smaller margin, in 3D.
Taken together, PQE and SPQE are very promising tools for studying many-body systems both in the strong and weak correlation regimes using NISQ hardware.

In summary, the PQE approach is a viable and more economical alternative to variational quantum algorithms.
In its current formulation, PQE can be applied to any trial state generated by exponentiating a set of linearly independent operators with identity metric matrix, as it is the case for disentangled unitary coupled-cluster ans\"{a}tze.
For these trial states, methods to reduce the number of measurements and exploit symmetries \cite{yen2020measuring, verteletskyi2020measurement, setia2020reducing} developed for VQE can likewise be used to improve PQE.
Interesting extensions of PQE include a generalization to unitaries that may contain repeated operators, that use generalized excitations/de-excitations pools, and hardware-efficient ans\"{a}tze.
In particular, a promising research direction is the formulation of a selected PQE using a basis of general one- and two-body operators, which could yield trial states with lower circuit depth than the current formulation.
Within the greater ecosystem of quantum algorithms, PQE could be used to determine initial guess amplitudes for subsequent optimization via VQE.
Additionally, similarly to VQE, PQE can be used as an alternative to adiabatic approaches to prepare initial states for quantum phase estimation.
Although we only explore applications of PQE for quantum many-body simulation, the framework outlined by Eqs.~\eqref{eq:ucc1} and \eqref{eq:ucc2} can be used to solve a variety of eigenvalue problems.
With appropriate modifications, for example, PQE could be employed to diagonalize covariance matrices (after quantum encoding \cite{giovannetti2008quantum, lloyd2014quantum}) for use in machine learning or data-analysis.
Moreover, because there is no requirement that the PQE (or SPQE) trial states have low-entanglement, PQE could be used as an alternative to methods such as quantum principle analysis \cite{lloyd2014quantum}, or variational quantum state diagonalization \cite{larose2019variational}.
Most importantly, the most immediate impact of PQE could be speeding up quantum computations on current or near-term devices.

\section*{Acknowledgements}
This work was supported by the U.S. Department of Energy under Award No.  DE-SC0019374 and the NSF under grant CHE-2038019.
N.H.S. was supported by a fellowship from The Molecular Sciences Software Institute under NSF grant ACI-1547580.

\appendix

\section{Gradient of the PQE energy}
\label{sec:appD}

In this section, we derive an expression for the PQE energy gradient. This result permits us to establish the equivalence of PQE and VQE for an exact UCC or dUCC ansatz, and to characterize the gradient when PQE is used to optimize approximate trial states.
Consider the PQE energy expression using a UCC [Eq.~\eqref{eq:trad_ucc}] or dUCC [Eq.~\eqref{eq:qucc}] ansatz
$E_\mathrm{PQE}(\mathbf{t}) = \bra{\Phi_0} \hat{U}^\dagger(\mathbf{t}) \hat{H} \hat{U}(\mathbf{t}) \ket{\Phi_0}$.
The derivative of $E_\mathrm{PQE}(\mathbf{t})$ with respect to the parameter $t_\mu$ is given by
\begin{equation}
\label{eq:appA:1}
\begin{split}
\frac{\partial E_\mathrm{PQE}(\mathbf{t})}{\partial t_\mu} & =
2 \, \mathrm{Re} \bra{\Phi_0}  \hat{U}^\dagger(\mathbf{t}) \hat{H} \frac{\partial \hat{U}(\mathbf{t})}{\partial t_\mu}  \ket{\Phi_0} \\
& = 2 \, \mathrm{Re} \bra{\Phi_0} \bar{H} \hat{U}^\dagger(\mathbf{t}) \frac{\partial \hat{U}(\mathbf{t})}{\partial t_\mu}  \ket{\Phi_0},
\end{split}
\end{equation}
where we have inserted the identity $\hat{U}^\dagger(\mathbf{t})  \hat{U}(\mathbf{t}) = 1$ to rewrite this expression in terms of $\bar{H}$.

Next, we consider the resolution of the identity
\begin{equation}
1 = \ket{\Phi_0}\bra{\Phi_0}
+ \sum_{\Phi_\mu \in R} \ket{\Phi_\mu}\bra{\Phi_\mu}
+ \sum_{\Phi_\nu \in S} \ket{\Phi_\nu}
\bra{\Phi_\nu},
\end{equation}
where $R$ is the set of determinants for which the residual condition $r_\mu = 0$ is enforced, while $S$ contains those determinants for which $r_\nu$ may not be equal to zero.
After inserting this resolution of the identity into Eq.~\eqref{eq:appA:1} between $\bar{H}$ and $\hat{U}^\dagger(\mathbf{t})$ and simplifying the resulting expression, we may express the gradient as
\begin{equation}
\label{eq:appA:3}
\begin{split}
\frac{\partial E_\mathrm{PQE}(\mathbf{t})}{\partial t_\mu}
 = \; &
2 E_\mathrm{PQE} \, \mathrm{Re}
\bra{\Phi_0} \hat{U}^\dagger(\mathbf{t}) \frac{\partial \hat{U}(\mathbf{t})}{\partial t_\mu}  \ket{\Phi_0} \\
& +
2 \, \mathrm{Re} \sum_{\Phi_\mu \in R}  r_\mu^* \bra{\Phi_\mu} \hat{U}^\dagger(\mathbf{t}) \frac{\partial \hat{U}(\mathbf{t})}{\partial t_\mu}  \ket{\Phi_0} \\
& +
2 \, \mathrm{Re} \sum_{\Phi_\nu \in S}
r_\nu^*
\bra{\Phi_\nu} \hat{U}^\dagger(\mathbf{t}) \frac{\partial \hat{U}(\mathbf{t})}{\partial t_\mu}  \ket{\Phi_0}.
\end{split}
\end{equation}
The first term in this expression is null since from $\bra{\Phi_0} \hat{U}^\dagger(\mathbf{t}) \hat{U}(\mathbf{t})\ket{\Phi_0} = 1$ we can show that the coefficient that multiplies the energy is null:
\begin{equation}
\frac{\partial \bra{\Phi_0} \hat{U}^\dagger(\mathbf{t}) \hat{U}(\mathbf{t})\ket{\Phi_0}}{\partial t_\mu} =
2 \; \mathrm{Re}\bra{\Phi_0} \hat{U}^\dagger(\mathbf{t}) \frac{\partial \hat{U}(\mathbf{t})}{\partial t_\mu} \ket{\Phi_0} = 0.
\end{equation}
The second term in Eq.~\eqref{eq:appA:3} is null due to the residual condition.
Applying these two simplifications we arrive at the gradient expression [Eq.~\eqref{eq:grad_residual_connection}] reported in Sec.~\ref{sec:trad_dis_ucc}, containing only the last term of Eq.~\eqref{eq:appA:3}.
The term $\bra{\Phi_\nu} \hat{U}^\dagger(\mathbf{t}) \frac{\partial \hat{U}(\mathbf{t})}{\partial t_\mu}  \ket{\Phi_0}$ that multiplies the residual is generally non-null for both the UCC and dUCC ans\"{a}tze.
In the case of dUCC, the term corresponding to the derivative with respect to $i$-th amplitude $t_{\mu_i}$, is given by:
\begin{widetext}
\begin{equation}
\begin{split}
\bra{\Phi_\nu} \hat{U}^\dagger(\mathbf{t}) \frac{\partial \hat{U}(\mathbf{t})}{\partial t_{\mu_i}}  \ket{\Phi_0}
=
\bra{\Phi_\nu}
e^{-t_{\mu_{N_\mathrm{op}}} \hat{\kappa}_{\mu_{N_\mathrm{op}}}}
\cdots
e^{-t_{\mu_{i+1}} \hat{\kappa}_{\mu_{i+1}}}
\hat{\kappa}_{\mu_i}
e^{t_{\mu_{i+1}} \hat{\kappa}_{\mu_{i+1}}}
\cdots
e^{t_{\mu_{N_\mathrm{op}}} \hat{\kappa}_{\mu_{N_\mathrm{op}}}}\ket{\Phi_0}.
\end{split}
\end{equation}
\end{widetext}
The states to the left and right of the operator $\hat{\kappa}_{\mu_i}$ are general many-body states that potentially span the entire Hilbert space, and therefore, this quantity is generally non-null.
For completeness, we also report the same expression for the case of traditional UCC, which may be obtained using the derivative of the exponential map:
\begin{widetext}
\begin{equation}
\begin{split}
\bra{\Phi_\nu} \hat{U}^\dagger(\mathbf{t}) \frac{\partial \hat{U}(\mathbf{t})}{\partial t_{\mu_i}}  \ket{\Phi_0}
=
\int_0^1
dx
\bra{\Phi_\nu}
e^{(x - 1)\hat{\sigma}}
\hat{\kappa}_{\mu_i}
e^{(1-x)\hat{\sigma}}
\ket{\Phi_0}.
\end{split}
\end{equation}
\end{widetext}

\section{Derivation of the \qucc-PQE update equations}
\label{sec:appC}

In this section, we provide a derivation of the amplitude update equation [Eq.~\eqref{eq:fixed_point}] based on a quasi-Newton method for solving general nonlinear equations.
To find update equations that relate new amplitudes ($\mathbf{t}^{(n+1)}$) to the previous set ($\mathbf{t}^{(n)}$), we expand the residual equation [Eq.~\eqref{eq:ucc2}] as a Taylor series centered around the current amplitude vector displaced by an amount $\Delta \mathbf{t}^{(n+1)} = \mathbf{t}^{(n+1)} - \mathbf{t}^{(n)}$
\begin{equation}
\begin{split}
r_\mu(\mathbf{t}^{(n+1)})   = \; & r_\mu(\mathbf{t}^{(n)} + \Delta \mathbf{t}^{(n+1)})  \\
 = \; & r_\mu(\mathbf{t}^{(n)}) + \sum_\nu J_{\mu\nu}(\mathbf{t}^{(n)})  \Delta t_\nu^{(n+1)} + \ldots,
\end{split}
\end{equation}
where $J$ is the Jacobian matrix, defined as $J_{\mu\nu}(\mathbf{t}) = \partial r_\mu(\mathbf{t}) / \partial t_\nu$.
To avoid computing and inverting the Jacobian, we seek a diagonal approximation to $J$.
We first note that for both the conventional UCC and the \qucc ans\"{a}tze, the similarity transformed Hamiltonian expanded up to linear terms is of the form
\begin{equation}
\hat{U}^\dagger(\mathbf{t}) \hat{H} \hat{U}(\mathbf{t})
= \hat{H} + t_{\mu_i}  \sum_i [\hat{H}, \hat{\kappa}_{\mu_i}]
+ \mathcal{O}(|\textbf{t}|^2).
\label{eq:appC:truncated_hbar}
\end{equation}
We now invoke the M{\o}ller--Plesset partitioning of the Hamiltonian and assume that the spin orbitals are canonical (i.e., they diagonalize the Hartree--Fock operator), allowing us to write $\hat{H} = \hat{F}^{(0)} + \hat{V}^{(1)}$, where $\hat{F}^{(0)} = \sum_{p} \epsilon_p \cop{p} \aop{p} $ is a zeroth-order diagonal one-body operator, and $\hat{V}^{(1)}$ is a first-order operator that contains two-body terms.
Approximating the Jacobian with with the first two terms of Eq.~\eqref{eq:appC:truncated_hbar} and retaining only the zeroth-order contributions, we obtain the following diagonal approximation
\begin{equation}
J_{\mu\nu}(\mathbf{t}) = \bra{\Phi_\mu} [\hat{F}^{(0)} , \hat{\kappa}_{\nu}]\ket{\Phi_0} = - \Delta_\mu \delta_{\mu \nu},
\end{equation}
where $\Delta_\mu = \epsilon_i + \epsilon_j + \ldots -\epsilon_a -\epsilon_b \ldots$ is a standard M{\o}ller--Plesset denominator corresponding to the excitation operator $\hat{\kappa}_\mu = \cop{a} \cop{b} \cdots \aop{j}\aop{i} - \mathrm{h.c.}$.
Inserting the diagonal Jacobian in the expanded residual we obtain
\begin{equation}
r_\mu(\mathbf{t}^{(n+1)})
= r_\mu(\mathbf{t}^{(n)}) - \Delta_\mu  \Delta t_\mu^{(n+1)},
\end{equation}
which when solved for $r_\mu(\mathbf{t}^{(n+1)}) = 0$ yields the update equation [Eq.~\eqref{eq:fixed_point}].

\section{Additional numerical comparison of PQE and VQE}
\label{sec:add_num_vqe_compare}

\begin{figure*}[ht!]
\centering
\includegraphics[width=6.0in]{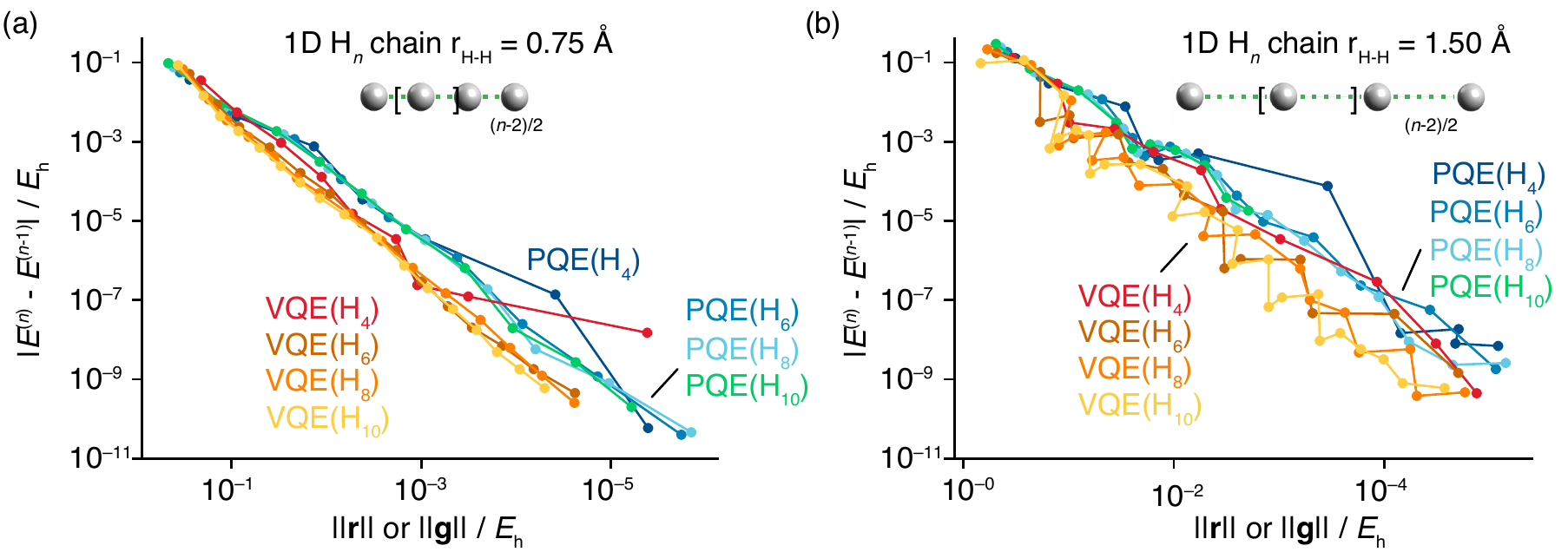}
\caption{dUCCSD energy convergence for linear \ce{H4}--\ce{H10} chains in a STO-6G basis at (a) $r_{\rm{H-H}} = 0.75$~{\AA}, and (b) $r_{\rm{H-H}} = 1.50$~{\AA}. $| E^{(n)} - E^{(n-1)}|$  is the absolute value of the energy change between subsequent iterations. Both plots compare PQE vs VQE convergence with respect to norm of the residual-vector (for PQE) or the norm of the gradient-vector (for VQE).}
\label{fig:Hn_econv_both_SI}
\end{figure*}

Here we provide additional numerical comparison for PQE and VQE. 
Table~\ref{tab:BeH2_Econv_comparison} shows numerics for ground state energy convergence of \ce{BeH2} using dUCCSD-PQE and dUCCSD-VQE. 
Figure~\ref{fig:Hn_econv_both_SI} shows the energy converge of various hydrogen chain systems with respect to the norm of the residual vector (for PQE) and gradient vector (for VQE).
For both Tab.~\ref{tab:BeH2_Econv_comparison} and Fig.~\ref{fig:Hn_econv_both_SI} trends similar to those seen in Fig.~\ref{fig:Hn_econv_both} are observed.

\begin{table*}[ht!]
\centering
\renewcommand{\arraystretch}{0.75}
\caption{Comparison of the convergence of the ground state energy of \ce{BeH2} computed with PQE and VQE using a disentangled UCCSD trial state. For a method $X$ = PQE or VQE, this table reports the total energy ($E_X$, in \Eh), the energy change from the previous to the current iteration ($E^{(n)}_X - E^{(n-1)}_X$), the number of residual/gradient evaluations ($N_X^{\cdot}$), and the norm of the residual/gradient ($\| \cdot \|$). The FCI energies at 1.0 and 2.0 \AA{} are $-15.65068726$ and $-15.60861964$ \Eh, respectively. All calculations use a STO-6G basis.}
\tiny 
\begin{tabular*}{\textwidth}{@{\extracolsep{\stretch{1.0}}}*{1}{c}*{8}{c}@{}}
    \hline

    \hline
   Iteration  &  $E_{\rm{PQE}}$  & $E^{(n)}_{\rm{PQE}} - E^{(n-1)}_{\rm{PQE}}$ & $N^{\rm{r-eval}}_{\rm{PQE}}$   & $\normnorm{\mathbf{r}_{\rm{PQE}}}$ & $E_{\rm{VQE}}$ & $E^{(n)}_{\rm{VQE}} - E^{(n-1)}_{\rm{VQE}}$ & $N^{\rm{g-eval}}_{\rm{VQE}}$  & $\normnorm{\mathbf{g}_{\rm{VQE}}}$   \\
    \hline
    
 \multicolumn{9}{c}{\ce{BeH2} ($r_\mathrm{Be-H}$ = 1.0 \AA) UCCSD}\\[6pt]
1    &    $-$15.6480381687   &   $-$0.0233010440    &     1          &        0.2184453066 &   $-$15.6448482964   &   $-$0.0201111716    &     3    &      0.3114581416    \\
2    &    $-$15.6500971948   &   $-$0.0020590260    &     2          &        0.0613327467 &   $-$15.6478816410   &   $-$0.0030333446   &     5    &      0.1481817759    \\
3    &    $-$15.6504069826   &   $-$0.0003097878    &     3          &        0.0088860708 &   $-$15.6494462890   &   $-$0.0015646480    &    7    &      0.1065575796    \\
4    &    $-$15.6504328168   &   $-$0.0000258343    &     4          &        0.0026511924  &   $-$15.6499421388  &   $-$0.0004958498    &     9    &     0.0634937794    \\
5    &    $-$15.6504349587   &   $-0.0000021418$   &     5          &       0.0005328227  &   $-$15.6500470587   &   $-$0.0001049199    &   10    &     0.0893255742    \\
6    &    $-$15.6504350041   &   $-$0.0000000454    &     6          &       0.0000763318  &   $-$15.6501955946   &   $-$0.0001485359    &   11    &     0.0657024179     \\

7    &    $-$15.6504350044   &   $-$0.0000000003    &     7          &       0.0000086198  &   $-$15.6502874046   &   $-$0.0000918100    &   12    &      0.0690985413    \\
8    &                                    &                                  &               &                               &   $-$15.6503399456   &   $-$0.0000525410    &     13    &      0.0453884569     \\
9    &                                    &                                  &               &                               &   $-$15.6503962522   &   $-$0.0000563066    &     14    &      0.0249260915     \\
10  &                                    &                                  &               &                               &   $-$15.6504228495   &   $-$0.0000265973    &     16    &      0.0137252110     \\
11  &                                    &                                  &               &                               &   $-$15.6504302548   &   $-$0.0000074053    &     18    &      0.0073189453     \\
12  &                                    &                                  &               &                               &   $-$15.6504341054   &   $-$0.0000038506    &     20    &      0.0037260965     \\
13  &                                    &                                  &               &                               &   $-$15.6504346315   &   $-$0.0000005261    &     22    &      0.0023579786     \\
14  &                                    &                                  &               &                               &   $-$15.6504348973   &   $-$0.0000002657    &     24    &      0.0014145610     \\
15  &                                    &                                  &               &                               &   $-$15.6504349615   &   $-$0.0000000642    &     26    &      0.0008586922     \\
16  &                                    &                                  &               &                               &   $-$15.6504349766   &   $-$0.0000000151    &     28    &      0.0006710048     \\
17  &                                    &                                  &               &                               &   $-$15.6504349959   &   $-$0.0000000193    &     29    &      0.0003518429     \\
18  &                                    &                                  &               &                               &   $-$15.6504350027   &   $-$0.0000000068    &     31    &      0.0001962939     \\
19  &                                    &                                  &               &                               &   $-$15.6504350038   &   $-$0.0000000011    &     33    &      0.0001228873     \\
20  &                                    &                                  &               &                               &   $-$15.6504350044   &   $-$0.0000000005    &     34    &      0.0000505365     \\
21  &                                    &                                  &               &                               &   $-$15.6504350045   &   $-$0.0000000001    &     36    &      0.0000433400     \\
22  &                                    &                                  &               &                               &   $-$15.6504350047   &   $-$0.0000000002    &     37    &      0.0000200500     \\[6pt]

 \multicolumn{9}{c}{\ce{BeH2} ($r_\mathrm{Be-H}$ = 2.0 \AA) UCCSD}\\[6pt]
 1    &    $-$15.5888292373   &   $-$0.0721251319    &     1         &         0.2551300809 &   $-$15.5797540529   &   $-$0.0630499476    &     3    &      0.7559489104    \\
 2    &    $-$15.6022316771   &   $-$0.0134024398    &     2         &         0.1072484292 &   $-$15.5993381767   &   $-$0.0195841238    &     5    &      0.2319720501    \\
 3    &    $-$15.6055237786   &   $-$0.0032921016    &     3         &         0.0175246100 &   $-$15.6027545167   &   $-$0.0034163400    &     7    &      0.1024944422    \\
 4    &    $-$15.6056880097   &   $-$0.0001642310    &     4         &         0.0076659595 &   $-$15.6041840035   &   $-$0.0014294868    &     9    &      0.1471217958    \\
 5    &    $-$15.6057527283   &   $-$0.0000647187    &     5         &         0.0051078803 &   $-$15.6050561955   &   $-$0.0008721920    &     10    &      0.0754837786    \\
 6    &    $-$15.6058044053   &   $-$0.0000516769    &     6         &         0.0012840244 &   $-$15.6052847256   &   $-$0.0002285301    &     12    &      0.0580899178    \\
 7    &    $-$15.6058068352   &   $-$0.0000024300    &     7         &         0.0002042506 &   $-$15.6055945458   &   $-$0.0003098202    &     13    &      0.0291205752    \\
 8    &    $-$15.6058068368   &   $-$0.0000000016    &     8         &         0.0000663613 &   $-$15.6056501930   &   $-$0.0000556471    &     15    &      0.0302968175    \\
 9    &    $-$15.6058068335   &   $+$0.0000000033   &      9        &         0.0000122091 &   $-$15.6057330619   &   $-$0.0000828689    &     16    &      0.0197063516    \\
10   &    $-$15.6058068336   &   $-$0.0000000000    &    10        &         0.0000040077 &   $-$15.6057681510   &   $-$0.0000350891    &     18    &      0.0186569595    \\
11   &                                     &                                    &                &                                 &   $-$15.6058033733   &   $-$0.0000352223    &     19    &      0.0069816557     \\
12   &                                     &                                    &                &                                 &   $-$15.6058052286   &   $-$0.0000018553    &     21    &      0.0045627712     \\
13   &                                     &                                    &                &                                 &   $-$15.6058057676   &   $-$0.0000005391    &     23    &      0.0036888638     \\
14   &                                     &                                    &                &                                 &   $-$15.6058065558   &   $-$0.0000007882    &     24    &      0.0021623560     \\
15   &                                     &                                    &                &                                 &   $-$15.6058067218   &   $-$0.0000001661    &     26    &      0.0015523193     \\
16   &                                     &                                    &                &                                 &   $-$15.6058068503   &   $-$0.0000001285    &     27    &      0.0003628532     \\
17   &                                     &                                    &                &                                 &   $-$15.6058068547   &   $-$0.0000000044    &     29    &      0.0002829316     \\
18   &                                     &                                    &                &                                 &   $-$15.6058068601   &   $-$0.0000000054    &     30    &      0.0000872433     \\
19   &                                     &                                    &                &                                 &   $-$15.6058068604   &   $-$0.0000000003    &     32    &      0.0000660842     \\
20   &                                     &                                    &                &                                 &   $-$15.6058068607   &   $-$0.0000000002    &     33    &      0.0000166869     \\[6pt]

    \hline
    \hline
\end{tabular*}
\label{tab:BeH2_Econv_comparison}
\end{table*}

\section{Formal comparison of PQE and VQE}
\label{sec:formal_vqe_compare}

We begin by assuming the case of a fixed ansatz, and since in this case state preparation costs are the same, our analysis focuses on how many measurements are necessary to compute the gradients and residuals in the iterative optimization procedure.

The preferred approach to minimize the energy in VQE is via gradient-based algorithms.
The gradient of the \qucc  energy with respect to a cluster amplitude $t_\mu$ is given by
\begin{equation}
\begin{split}
\frac{\partial E_\mathrm{VQE}(\mathbf{t})}{\partial t_\mu} = &
2 \, \mathrm{Re} \bra{\Phi_0}  \hat{U}^\dagger(\mathbf{t}) \hat{H} \frac{\partial \hat{U}(\mathbf{t})}{\partial t_\mu}  \ket{\Phi_0}
\label{eq:grad_mu_measurable}
\end{split}
\end{equation}
Equation~\eqref{eq:grad_mu_measurable} has the form of an off-diagonal matrix element of the Hamiltonian, and Romero et al. \cite{Romero:2019hk} showed that it may be measured on a quantum device using one ancilla qubit and four time the cost of measuring the energy.
With this algorithm, computing the gradient vector so that the 2-norm of the error is within precision $\bar{\epsilon}_{\rm{grad}}$ requires a total number of measurements ($m_\text{grad}$) that is bound by the inequality \cite{Romero:2019hk}
\begin{equation}
\label{eq:mgrad}
m_\text{grad} \leq 4 N_{\rm{par}} \frac{ \big( \sum_\ell | h_\ell | \big)^2 }{ \bar{\epsilon}_{\rm{grad}}^2  }.
\end{equation}

The total number of measurements, and proportionally the runtime, required for a VQE calculation is dominated by the computation of the gradient vector and it is proportional to the number of gradient vector evaluations ($N_{\rm{grad}}^{\rm{VQE}}$) required to converge the energy. This implies that the total number of VQE measurements ($m_{\rm{VQE}} $) is bounded approximately by
\begin{equation}
m_{\rm{VQE}} \leq N_{\rm{grad}}^{\rm{VQE}} m_\text{grad}.
\end{equation}
This estimate ignores the evaluation of the VQE energy, which has a cost inferior to that of computing one element of the gradient vector.
More recently, Kottmann et al. \cite{kottmann2020feasible} showed that the analytic gradient may be computed with a cost essentially equal to that of two energy evaluations using the so-called  parameter-shift-rule \cite{schuld2019evaluating}.
This procedure avoids the use of an ancilla qubit and the number of measurements required still satisfies the bound expressed in Eq.~\eqref{eq:mgrad}.

In the case of PQE, the number of measurements $m_\text{res}$ needed to compute the residual vector with precision $\bar{\epsilon}_{\rm{res}}$ has an upper bound given by
\begin{equation}
m_\text{res} \leq 3 N_{\rm{par}} \frac{ \big( \sum_\ell | h_\ell | \big)^2 }{ \bar{\epsilon}_{\rm{res}}^2 }.
\end{equation}
This estimate takes into account the fact that the residual can be evaluated as the sum of three terms (with different prefactors), and that $E_0$ in Eq.~\eqref{eq:res_measure} only needs to be measured once.
The total number of PQE measurements ($m_\text{PQE}$) is bounded by
\begin{equation}
m_\text{PQE} \leq N_\text{res}^\text{PQE} m_\text{res},
\end{equation}
where $N_\text{res}^\text{PQE}$ is the number of PQE residual vector evaluations.
Assuming that the energy gradients in VQE and the residual in PQE are measured with the same precision ($\bar{\epsilon}_\text{grad} \approx \bar{\epsilon}_{\text{res}}$), we estimate that  $m_\text{res}  \approx \frac{3}{4} m_\text{grad}$.
This result suggest that PQE should have a similar or perhaps slightly smaller cost per iteration than VQE.
However, a more important factor in determining the relative performance of VQE and PQE is the number of residual/gradient evaluations required, which as  shown in Sec.~\ref{sec:results}, favors PQE over VQE.

A detailed comparison of the adaptive variants of VQE (e.g., ADAPT-VQE) and PQE is more complex due to the significant differences in the form of the ansatz and the selection procedure used in these two methods.
At iteration $k$, the ADAPT-VQE approach selects the operator $\hat{\kappa}_\mu$ with the largest absolute energy gradient.
This selection scheme requires evaluating the gradient $g_\mu^{(k)}$ for all the operators in the pool
\begin{equation}
g_\mu^{(k)} = \bra{\Psi^{(k)}} [ \hat{H}, \hat{\kappa}_\mu ] \ket{\Psi^{(k)}}.
\label{eq:adapt_imp_metric}
\end{equation}
Because most sub-terms of the Hamiltonian will commute with $\hat{\kappa}_\mu$, a relatively small number of $\braket{ \hat{\kappa}_\mu \hat{O}_\ell }$ and $\braket{ \hat{O}_\ell \hat{\kappa}_\mu }$ terms need to be measured.
Both $\hat{H}$ and the operator pool $\{ \hat{\kappa}_\mu \}$ contain of the order $N^4$ elements (assuming a pool of general one and two-body operators).

However, it has recently been pointed out \cite{liu2020efficient} that if the quantity $[ \hat{H}, \hat{\kappa}_\mu ]$ is decomposed in terms of the reduced density matrices, then one can determine the pool gradients [Eq.~\eqref{eq:adapt_imp_metric}] with the evaluation of a number of Pauli terms that scales as $N^6$ (again assuming a pool of general one and two-body operators).
In ADAPT-VQE this cost must then be multiplied by the number of iterations performed.
It is important to note that the ADAPT-VQE macro-iteration convergence threshold $\epsilon_\alpha = 10^{-\alpha}$ is based on the norm of the vector of pool gradients [Eq.~\eqref{eq:adapt_imp_metric}], such that ADAPT-VQE is considered converged when $\normnorm{\mathbf{g}_{\rm{pool}}}  \leq \epsilon_\alpha$.
The number of measurements of the approximate residual vector ($\ket{\tilde{r}}$) for the purpose of selection in SPQE is a parameter of a computation. In Appendix~\ref{apdx:aa_reduce_cost}, we show that a probabilistic estimate for the number of measurements required to converge SPQE with a threshold of $\Omega$ is of the order of $(\Delta t \Omega )^{-2}$.

A  trade-off of using three- and higher-body operators in SPQE is a greater circuit depth since an operator $\hat{\kappa}_{ij\cdots}^{ab\cdots}$ of many-body rank $n$ becomes  a sum of $2^{2n -1}$ Pauli strings after Jordan--Wigner mapping to the qubit basis.
Since all Pauli strings that are generated in this mapping commute, the unitary $\exp(\theta \, \hat{\kappa}_{ij\cdots}^{ab\cdots})$ can be written as the product of $2^{2n -1}$ exponentials of operators containing Pauli strings of length $2n$.

To analyze the compromise between the rank of the operator pool and the compactness of the ansatz in ADAPT-VQE and SPQE we consider the example of a three-body operator $\exp(t \, \hat{\kappa}_{ijk}^{abc})$.
In ADAPT-VQE this term may be approximated using general two-body operators as
\begin{equation}
\label{eq:3body_example}
\begin{split}
e^{t \, \hat{\kappa}_{ijk}^{abc}}
=
e^{t \, [\hat{\kappa}_{ij}^{ae},\hat{\kappa}_{ke}^{bc}]}
\approx
e^{t \, \hat{\kappa}_{ij}^{ae}}
e^{t \, \hat{\kappa}_{ke}^{bc}}
e^{-t \,\hat{\kappa}_{ij}^{ae}}
e^{-t \, \hat{\kappa}_{ke}^{bc}},
\end{split}
\end{equation}
where the last term is a lowest-order Trotter approximation of the exponential of the commutator $[\hat{\kappa}_{ij}^{ae},\hat{\kappa}_{ke}^{bc}]$ (with $e \neq a,b,c$).
The last term in Eq.~\eqref{eq:3body_example} is implemented as a circuit that contains four different parameters and a product of 32 exponentials of Pauli strings of length four.
The same three-body excitation is represented in SPQE using a single parameter and a longer circuit as a product of 32 exponentials of Pauli strings of length six.
This comparison suggests, in accordance with the results of our study, that higher-body operators are represented less efficiently with an arbitrary particle-hole operator pool than a general singles and doubles operator pool.

\section{Reduced-cost estimation of the approximate residual in selected PQE}
\label{apdx:aa_reduce_cost}

This appendix explores various methods to reduce the number of measurements $M$ required to compute the approximate the residuals $\tilde{r}_\mu$ used in the selected PQE method (see  Sec.~\ref{sec:selection}).
Selection requires the identification of the elements of the approximate residual that corresponds to projections onto excited determinants. However, for small values of $\Delta t$, the state $\ket{\tilde{r}}$ is dominated by the reference determinant $\Phi_0$, and consequently, the measurement of important missing excitations may become inefficient.
\begin{figure}[h!]
\centering
\includegraphics[width=3.375in]{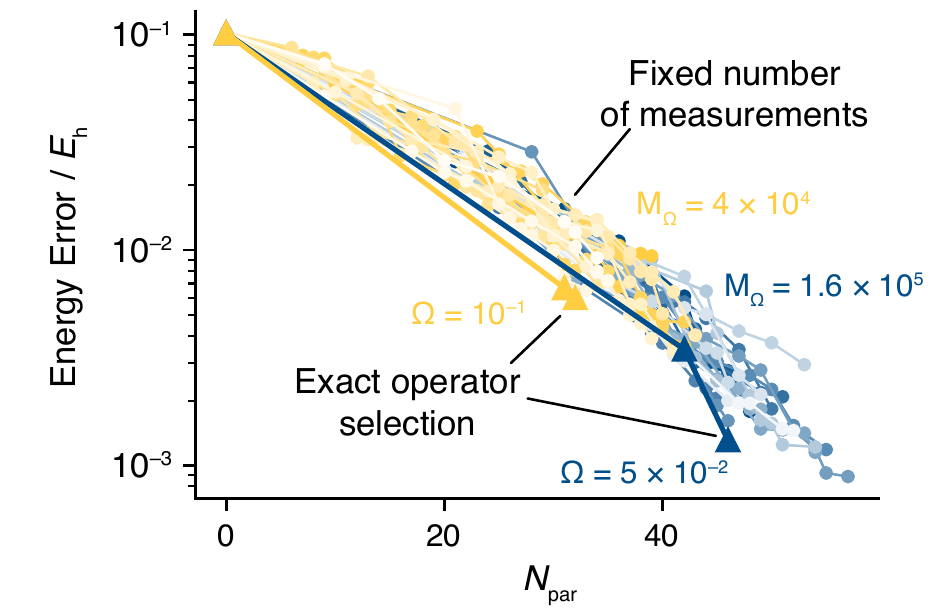}
\caption{SPQE energy convergence for \ce{H6} at a separation of $r_{\rm{H-H}} = 1.0$~\AA, with $\Delta t = 0.05$ au. Data sets with triangular markers denote exact operator selection and convergence via Eqs.~\eqref{eq:approx_res_sq} and~\eqref{eq:op_thresh} for $\Omega = 1.0\times10^{-1}$ (yellow) and $\Omega = 5.0\times10^{-2}$ (dark blue). Data sets with dots denote operator selection and convergence with a fixed number of measurements $M_{\Omega} = 4.0\times10^{4}$ (yellow) and $M_{\Omega} = 1.6\times10^{5}$ (dark blue) calculated via Eq.~\eqref{eq:fixed_measure_thresh} for the corresponding $\Omega$ values.}
\label{fig:Momega}
\end{figure}
In practice one can consider an alternative convergence criterion for SPQE based on performing a fixed number of measurements $M_{\Omega}$ on the state $\ket{\tilde{r}}$.
In such an approach, at each iteration $k$, the operators $\hat{\kappa}_\mu$ whose corresponding determinants $\ket{\Phi_\mu}$ are measured at least once (over all $M_{\Omega}$ measurements) are added to $\mathcal{A}$.
Because the residual magnitudes go to zero as the eigenstate is better approximated in successive $k$ iterations, it becomes increasingly unlikely that any determinants besides the reference $\ket{ \Phi_0 }$ will be measured.
The SPQE algorithm can then be considered converged when all $M_{\Omega}$ measurements yield the reference state.
Starting from Eqs.~\eqref{eq:approx_res_sq} and~\eqref{eq:op_thresh}, and making the assumption that at convergence only a single determinant $\Phi_\mu$ is measured that is not the reference (i.e., $\sum_\mu N_\mu = 1$), one can use a number of measurements
\begin{equation}
\label{eq:fixed_measure_thresh}
M_\Omega = \frac{1}{\Delta t^2 \Omega^2},
\end{equation}
to probabilistically test convergence of the residual vector to within the threshold $\Omega$.
In practice we find that using Eq.~\eqref{eq:fixed_measure_thresh} works well compared to the exact threshold given in Eq.~\eqref{eq:op_thresh}.
Figure~\ref{fig:Momega} compares the energy convergence with number of selected operators/parameters using both convergence criterion corresponding to the same value of $\Omega$.

\bibliography{bibliography}

\end{document}